\documentclass[journal]{IEEEtran}
\usepackage[T1]{fontenc}
\usepackage{float}
\usepackage{amsmath}
\usepackage{amsthm}
\usepackage{amssymb}
\usepackage{stackrel}
\usepackage{graphicx}
\usepackage{setspace}
\usepackage{url}
\allowdisplaybreaks

\usepackage[utf8]{inputenc}

\usepackage{caption}
\usepackage{subfigure}
\usepackage{algorithm}
\usepackage[graphicx]{realboxes}
\usepackage{algorithmic}
\usepackage{verbatim}

\makeatletter

\floatstyle{ruled}
\newfloat{algorithm}{tbp}{loa}
\providecommand{\algorithmname}{Algorithm}
\floatname{algorithm}{\protect\algorithmname}
\theoremstyle{plain}

\theoremstyle{definition}

\theoremstyle{plain}

\theoremstyle{plain}
\newcommand{\RNum}[1]{\uppercase\expandafter{\romannumeral #1\relax}}
\usepackage[thicklines]{cancel}
\IEEEoverridecommandlockouts
\usepackage{ifpdf}
\usepackage{cite}
\usepackage{graphicx}
\ifCLASSOPTIONcompsoc
\else
\fi
\usepackage{graphicx}
\usepackage{epstopdf}
\usepackage{amsmath}
\usepackage{mathtools}
\usepackage{booktabs} 
\usepackage{multirow}
\usepackage{setspace}
\usepackage{lettrine} 
\usepackage{bm}
\makeatother
\usepackage[]{babel}

\newtheorem{remark}{Remark}

\usepackage{booktabs} % 导入三线表需要的宏包
\usepackage{longtable}% 导入跨页表格所需宏包
\usepackage{amsthm} % 如果要使用proof语句就必须要引入这个包
\usepackage{stfloats}
\usepackage{colortbl}  %彩色表格需要加载的宏包
\usepackage{xcolor}
\usepackage{array}   %对表列和表格线的设置需要用到array宏包

\begin{document}
	\captionsetup[figure]{font={small}, name={Fig.}, labelsep=period}
	\title{Lightweight and Self-Evolving Channel Twinning: An Ensemble DMD-Assisted Approach}
	\author{
		Yashuai Cao,~\IEEEmembership{Member,~IEEE},
		Jintao Wang,~\IEEEmembership{Senior Member,~IEEE},\\
		Xu Shi,~\IEEEmembership{Graduate Student Member,~IEEE},
		and Wei Ni,~\IEEEmembership{Fellow,~IEEE}
		\thanks{This work of Yashuai Cao was supported in part by the Project funded by China Postdoctoral Science Foundation under Grant No. 2023M742009; in part by the Postdoctoral Fellowship Program of CPSF under Grant No. GZC20231372; in part by the Project funded by China Postdoctoral Science Foundation under Grant 2024M761671; and in part by Beijing Natural Science Foundation under Grant 4254069.}
		\thanks{Yashuai Cao, Jintao Wang, and Xu Shi are with the Department of Electronic Engineering, Tsinghua University, Beijing 100084, China, and also with the Beijing National Research Center for Information Science and Technology (BNRist), Beijing 100084, China (e-mail: \{caoys, wangjintao, shi-x\}@tsinghua.edu.cn.}
		\thanks{Wei Ni is with CSIRO, Sydney, New South Wales, 2122, Australia (e-mail: wei.ni@data61.csiro.au).}
	}
	
	\maketitle
	\begin{abstract}
		Traditional channel acquisition faces significant limitations due to ideal model assumptions and scalability challenges. A novel environment-aware paradigm, known as channel twinning, tackles these issues by constructing radio propagation environment semantics using a data-driven approach. In the spotlight of channel twinning technology, a radio map is recognized as an effective region-specific model for learning the spatial distribution of channel information. However, most studies focus on static channel map construction, with only a few collecting numerous channel samples and using deep learning for radio map prediction. 
		In this paper, we develop a novel dynamic radio map twinning framework with a substantially small dataset. Specifically, we present an innovative approach that employs dynamic mode decomposition (DMD) to model the evolution of the dynamic channel gain map as a dynamical system. We first interpret dynamic channel gain maps as spatio-temporal video stream data. The coarse-grained and fine-grained evolving modes are extracted from the stream data using a new ensemble DMD (Ens-DMD) algorithm. To mitigate the impact of noisy data, we design a median-based threshold mask technique to filter the noise artifacts of the twin maps. With the proposed DMD-based radio map twinning framework, numerical results are provided to demonstrate the low-complexity reproduction and evolution of the channel gain maps. Furthermore, we consider four radio map twin performance metrics to confirm the superiority of our framework compared to the baselines.
	\end{abstract}
	
	
	\section{Introduction}\label{sec:1}
	\IEEEPARstart{C}{hannel} twinning has gained increasing interest recently, with the prevalence of massive multiple-input-multiple-output (MIMO) and high-mobility communication scenarios in the sixth-generation (6G) mobile networks~\cite{10375691, cao2024channel, 10742564}. By integrating various state observations, such as three-dimensional (3D) maps, historical position estimates, and channel state information (CSI) measurements, wireless network operations can be performed efficiently according to the emulated CSI from the channel twinning model~\cite{3652190}.
	Channel twinning technology breaks through the limitations of mathematical models by fully leveraging environmental information. It can significantly reduce the channel acquisition overhead while ensuring prediction precision. To this end, 3D environment models can be obtained directly from OpenStreetMap (OSM), but are usually combined with Blender in channel twinning technology for fast ray tracing~\cite{borges2024caviar}.
	
	To explore the benefits of channel twinning, radio map technology, also known as channel knowledge map~\cite{9373011, 10352759}, has been viewed as an enabler toward environmental awareness. A radio map is a region-specific model that can help obtain channel-related information about any locations in the region of interest.
	In radio networks, radio maps offer vital information for resource allocation tasks, eliminating the need for channel estimation pilots~\cite{10430216}.
	In practice, mobile devices (MDs) experience non-static channel variations due to the Doppler shift. Accordingly, static radio maps cannot capture the time evolution of channels, as the real-time emulation of MDs' dynamic behavior is extremely challenging.
	One of the most urgent issues for channel twinning is how to handle the dynamic twin efficiently.

	\subsection{Channel Evolution Analysis and Challenge}\label{sec:1.1}
	To model the channel dynamics, the Doppler shift model is widely applied in traditional studies. By only considering the time-varying Doppler effect on the current CSI instead of electromagnetic (EM) environmental dynamics, the Doppler shift model-based channel evolution is given by~\cite{8580850}
	\begin{align}
		\hat{h}(f, t) &= \sum_{l=0}^{L-1} \alpha_l \mathrm{e}^{j2\pi f_{\mathrm{d}} t} \mathrm{e}^{-j2\pi f \tau_{l}}, \label{eq:1}
	\end{align}
	where $L$ is the number of channel paths, $\alpha_l$ is the initial path coefficient, $\tau_l$ is the path delay, and $f$ is the carrier frequency. Doppler shift $f_{\mathrm{d}}$ can be computed according to velocity vectors and ray directions of scattering points.
	
\begin{figure}[t]
		\centering{}\includegraphics[width=3.5in]{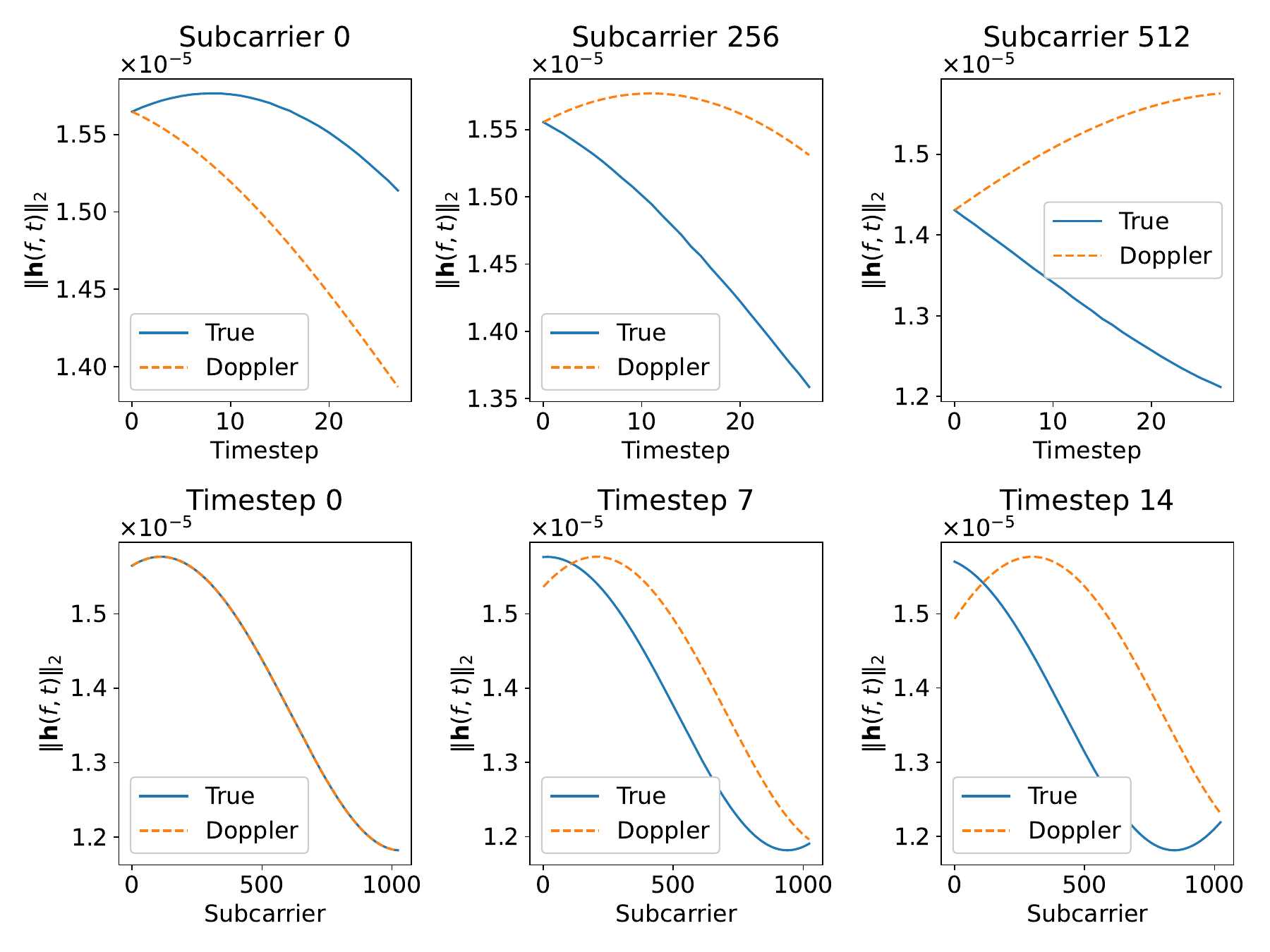}
		\caption{Comparison of Doppler shift model-based and physical environment-aware true channel evolution. The channel between a static transmitter and a mobile receiver is simulated in a simple street scene loaded by Sionna~\cite{10465179}. The transmitter is equipped with an $8\times 4$ uniform planar array, while the receiver utilizes a single antenna. The receiver's velocity is set to 30 m/s. The carrier frequency is 28~GHz. Each OFDM resource grid comprises $14 \times 2$ symbols, which correspond to 1024 subcarriers with a subcarrier spacing of 30 kHz.}
		\label{fig:1}
	\end{figure}

	To study the channel evolution precision, real CSI is acquired on the Sionna platform~\cite{10465179} by simulating the mobility of a receiver within a dynamic street environment. According to the Doppler shift model in~\eqref{eq:1}, the future CSI can be derived from the known initial channel coefficients, as shown in Fig.~\ref{fig:1}. 
	Compared to the true CSI, the Doppler-based channel evolution becomes noticeably less accurate as the number of time slots increases. The Doppler models ignore the dynamic changes in both ray directions and the number of propagation paths as the receiver moves.
	Naturally, applying the Doppler models directly to the static radio map cannot match the channel in realistic environments~\cite{7542162}.
	Unfortunately, existing channel emulation platforms, e.g.,~\cite{10465179, 9504428, 10634489}, are unable to capture instantaneous channel variations in highly dynamic environments.
	Thus, the identification of an effective dynamic twin modeling approach constitutes a significant challenge for the practical implementation of radio map technology.
	
	% \begin{align}
		% f_\Delta = \frac{\mathbf{v}_{0}^\mathsf{T}\hat{\mathbf{k}}_0 - \mathbf{v}_{n+1}^\mathsf{T}\hat{\mathbf{k}}_n + \sum_{i=1}^n \mathbf{v}_{i}^\mathsf{T}\left(\hat{\mathbf{k}}_i-\hat{\mathbf{k}}_{i-1} \right)}{\lambda}.
		% \end{align}
	
	\subsection{Related Studies}\label{sec:1.2}
\begin{figure}[t]
		\centering{}\includegraphics[width=3.5in]{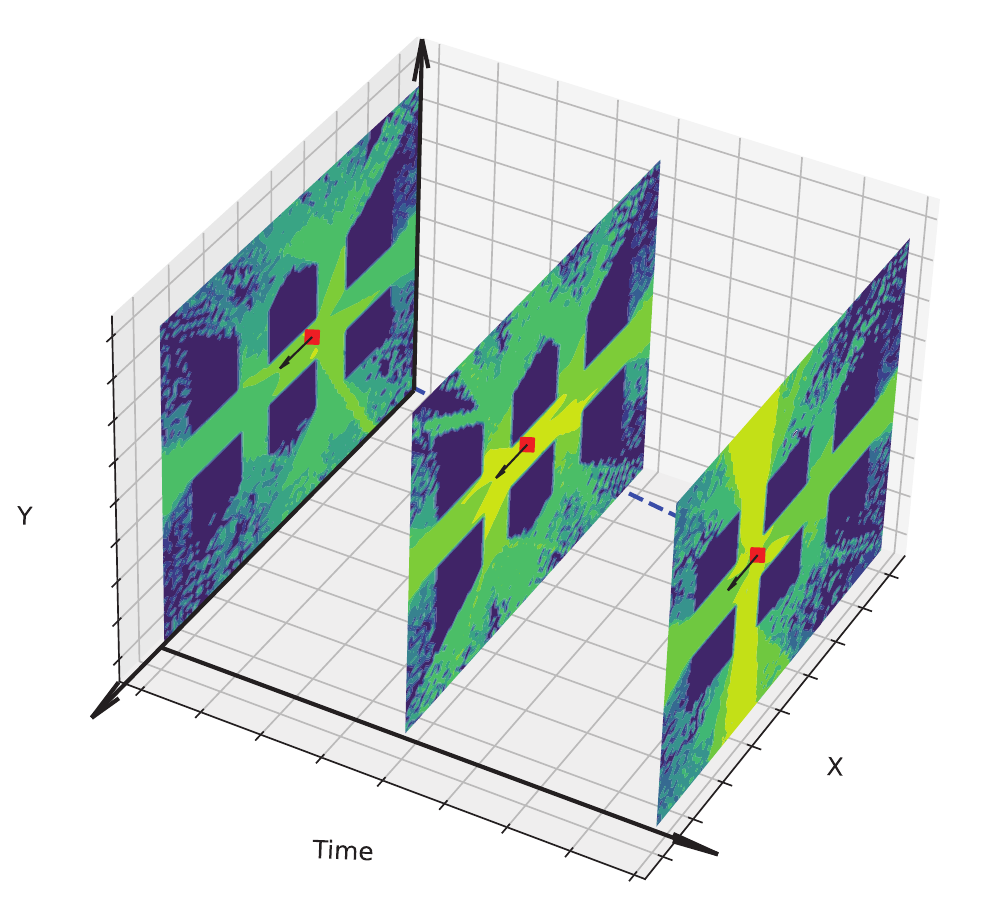}
		\caption{Dynamic variations of channel gain over time with an MD (indicated by a red marker) moving at a speed of 50 m/s.}
		\label{fig:2}
	\end{figure}

	The promising advancements in radio map technology for resource management in modern wireless communication systems have been revealed~\cite{8648450}. 
	A key type of radio map is the channel gain map, which records the spatial distribution of channel gain values based on the fixed location of a base station (BS). 
	In fifth-generation (5G) industrial environments, where the BS may relocate every few days or weeks to monitor production, the critical importance of dynamic channel gain maps is highlighted in~\cite{9310285, 10447973}.
	In addition to the need for industrial 5G networks, dynamic radio map twinning has recently gained significant interest in vehicle-to-everything (V2X) communications~\cite{10292913}. Dynamic and reliable intra-platoon communications require low-latency radio map querying and precise blockage prediction.
	In such applications, dynamic spectrum management with moving vehicle radiation sources recognizes the importance of time-varying radio map prediction~\cite{cmu212560}. Fig.~\ref{fig:2} illustrates the temporal evolution of channel gain maps in a street scenario, where a vehicle-mounted radiator (i.e., MD) moves. 
	The coverage region of interest is discretized into grids, where each grid corresponds to a specific channel gain value. The channel gains across all grids form a radio map. Each frame of the radio map remains stable over short time intervals. The map frames are arranged in the time domain to reveal the radio environment dynamics, as shown in Fig.~\ref{fig:2}. These maps are typically collected by fixed or mobile monitoring nodes. Since it is impractical to collect maps in real time, the ability to predict future map states based on historical frame sequences emerges as a prerequisite.

	To realize dynamic radio maps, studies~\cite{5484600, 5711699} investigated the spatio-temporal evolution of channel gains in cognitive radio networks, specifically targeting the tracking of time-varying shadow fading. To this end, a Kriged Kalman filtering (KKF) algorithm~\cite{9931518, 10430216} was designed to construct dynamic channel gain maps. KKF typically requires the inverse of covariance matrices, which can become very expensive when the dimensionality of the state variables is high.
	Further, deep learning (DL) methods were introduced for radio map estimation~\cite{9354041, wang2024radiodiff}, which require a large dataset of radio maps. However, channel measurement activities are costly and time-consuming. Hence, the radio map dataset comprises few samples, rendering DL methods ineffective.

	Dynamic mode decomposition (DMD), as a popular tool for modeling complex dynamical systems,  has received little attention and application in the research of wireless communications until now. However, the potential of DMD for wireless communications has yet to be explored, with only two initial research efforts~\cite{10453222, 10279471} emerging.
	Therein, DMD methods are applied to infer the CSI generated by first-order auto-regressive (AR) models, focusing solely on tracking the CSI of a single user without considering the influence of the real-world environment.
	
	A series of powerful DMD variants, such as compressed DMD (cDMD)~\cite{kutz2016dynamic}, extended DMD (eDMD)~\cite{9147729}, and forward-backward DMD (fbDMD)~\cite{10188588}, remain within the scope of studying fluid flow control systems.
	DMD is a data-driven model order reduction technique that does not rely on system equations and can directly extract dynamic features from data~\cite{kutz2016dynamic}.
	Notably, DMD provides an equation-free framework that effectively describes the evolution processes of dynamic systems, with excellent interpretability and predictability.
	Due to its appealing processing ability of nonlinear and high-dimensional sequential data, DMD is recognized as a well-suited solution for implementing data-driven digital twin models~\cite{2023127317, 10221774}.

	\subsection{Contributions}\label{sec:1.3}
	In this study, we develop a new dynamic radio map twinning framework, where an ensemble DMD (Ens-DMD) algorithm is proposed to emulate the evolution of channel knowledge maps. The main contributions of the study are outlined as follows:%%小样本
	\begin{itemize}
		\item Inspired by the ability of DMD to predict the dynamics of complex fluid dynamics systems, we model the evolution of a dynamic channel gain map as a dynamical system. Based on the dataset collected from the Sionna platform, we treat the changing channel gain map as video stream data and develop a DMD-based twin framework for the reproduction and evolution of the channel gain map.
		\item Applying the standard DMD method directly struggles to capture local transient phenomena caused by mobility when twinning the channel gain map.
        Standard DMD tends to treat these transient dynamics as noise~\cite{kutz2016dynamic}. To address this limitation, we propose an eDMD-based channel gain map twin modeling approach. By leveraging kernel functions, eDMD maps the measurement data into a high-dimensional feature space, allowing for the effective capture of highly nonlinear transient dynamics.
		\item In the presence of noisy measurements, the eDMD method may misinterpret noise as nonlinear features, compromising twin accuracy. To address this challenge, we propose an Ens-DMD algorithm that incorporates a median threshold mask technique to separately process the outputs of both eDMD and cDMD methods. This effectively filters out noise while retaining critical transient features. 
		\item 
		We further integrate the Ens-DMD process with a Kriging interpolation algorithm to develop a dynamic radio map twinning framework. We evaluate the framework using several performance metrics for the radio map and compare the proposed method against the benchmark algorithms, demonstrating its superior efficiency and robustness.
	\end{itemize}
	
	The structure of this paper is organized as follows. Section~\ref{sec:2} describes the dynamic radio map twinning problem. In Section~\ref{sec:3}, we provide the fundamentals of DMD and its variants, present the proposed Ens-DMD algorithm, and outline the overall dynamic radio map twinning framework, along with an analysis of the twin modeling complexity. Section~\ref{sec:4} discusses the twin performance metrics, followed by simulation results that evaluate the performance of the radio map twin method.
	Section~\ref{sec:5} concludes the conclusions.
	
	\emph{Notation}: various lower- and upper-case boldface letters indicate column-vectors and matrices, respectively; $(\cdot)^{\mathsf T}$, $(\cdot)^{\mathsf H}$, $(\cdot)^{-1}$, and $(\cdot)^\dagger$ denote the transpose, conjugate transpose, inverse, and pseudo-inverse operations, respectively; $\mathrm{diag}(\mathbf{a})$ returns a square diagonal matrix with the vector $\mathbf{a}$ on the main diagonal; $\mathrm{Median}(\cdot)$ returns the median of a given set of numbers; $\mathbb{E}\{\cdot\}$ denotes the expectation; $\Vert \mathbf{a} \Vert_2$ denotes the Euclidean norm of a vector $\mathbf{a}$; $\Vert \mathbf{A} \Vert_F$ denotes the Frobenius norm of a matrix $\mathbf{A}$; and $\odot$ signifies the Hadamard product.
	
	\begin{figure}[t]
		\centering
		\subfigure[]
		{\includegraphics[width=3.5in]{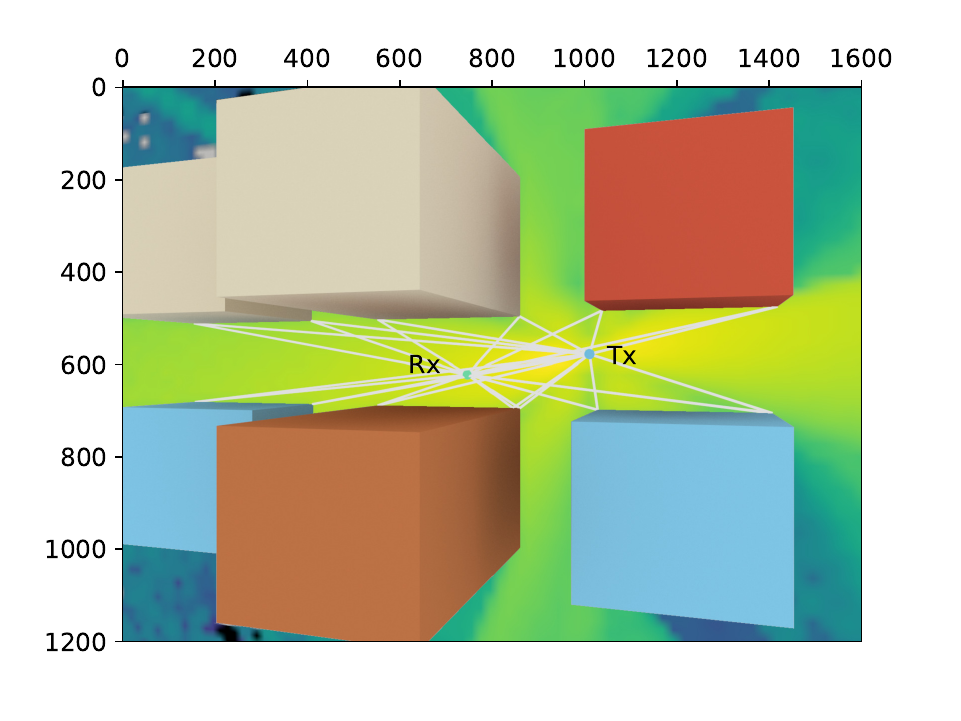}} 
		\subfigure[]
		{\includegraphics[width=3.5in]{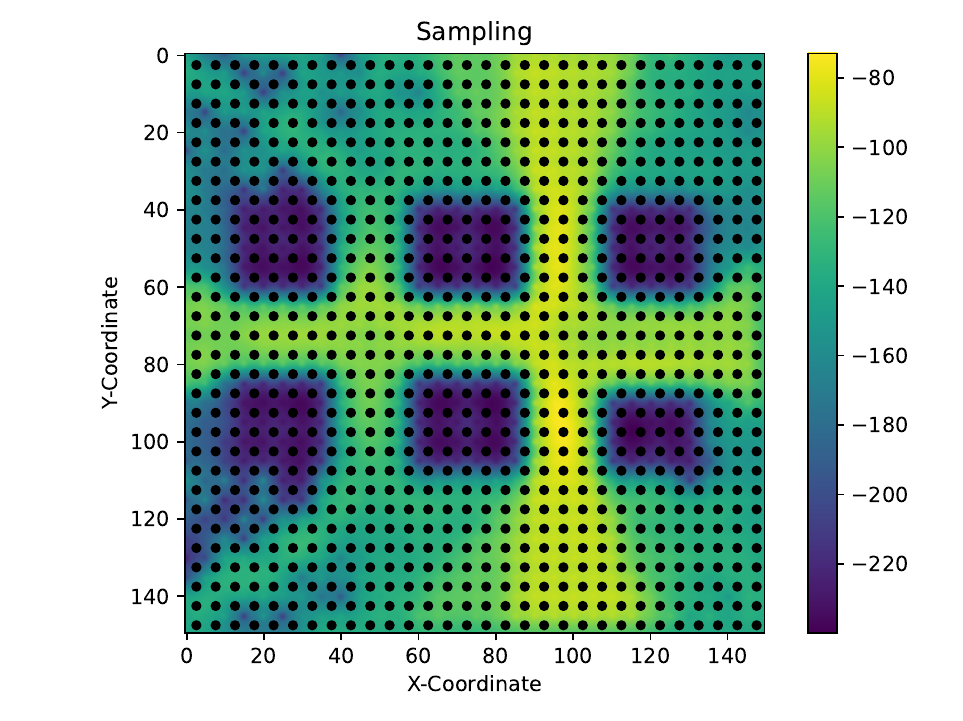}}
		\caption{Conceptual illustration of radio map construction: (a) An example scene of a simple street from Sionna~\cite{10465179}, where the receiver records radio measurements at various locations, and (b) a collection of measurements obtained from grid sampling points, serving as foundational data for the KKF.}
		\label{fig:3}
	\end{figure}
	
	\section{Dynamic Radio Map Twinning Problem}\label{sec:2}
	Traditional static channel knowledge maps operate under the assumption that the surrounding environment (such as the locations of BSs and nearby buildings) remains constant~\cite{9310285}. However, in 6G applications, the technology is challenged by dynamic environments. Consequently, radio resource allocation strategies based on static channel knowledge maps may experience reduced network performance. 
	As channel properties fluctuate over time due to environmental changes, the radio map can rapidly become outdated. Furthermore, frequent data calibration may be impractical or costly. To this end, this section explores the dynamic radio map construction problem through twin modeling based on historical channel measurements, while also considering the potential impact of measurement noise on the collected data.
	The core of the studied problem is to extract the channel evolution rules from a limited number of historical channel measurement samples, thereby constructing a radio map twinning model without the need for frequent collection and calibration. This twin model allows for the prediction or reconstruction of the radio map at any given moment.

	%\subsection{Radio Map Construction}\label{sec:2.1}
	Consider a geographical region $\mathcal{D}\in\mathbb{R}^{2}$ covered by a transmitter emitting radio frequency signals at the point $\mathbf{q}_0$, as shown in Fig.~\ref{fig:3}(a).
	The channel link gain from position $\mathbf{q}_0$ to $\mathbf{q}_i$ is expressed in decibels (dB) as:
	\begin{align}
		G_{ \mathbf{q}_0 \rightarrow \mathbf{q}_i} = G_{0} - 10 \gamma\log_{10} \left(\Vert \mathbf {q}_0 - \mathbf {q}_i \Vert\right) \nonumber\\
		+ s_{\mathbf{q}_0 \rightarrow \mathbf{q}_i} + \nu_{\mathbf{q}_0\rightarrow \mathbf{q}_i},
		\label{eq:2}
	\end{align}
	where $G_{0}$ is the path gain at a unit distance, $\gamma$ is the path-loss exponent, $\Vert{\cdot}\Vert$ is the Euclidean norm, $s_{\mathbf{q}_0 \rightarrow \mathbf{q}_i}$ is the shadow fading, and $\nu_{\mathbf{q}_0 \rightarrow \mathbf{q}_i}$ is the small-scale fading.

	%% y离散取值 
	%% s+w+noise
	%% 克里金
	To construct the radio map, channel measurements from network users at discrete locations, accompanied by location tags~\cite{xu2023much}, are collected, as shown in Fig.\ref{fig:3}(b). Assuming $M$ measurements taken from $\mathbf{q}_i \in \mathcal{D}$, we denote the set of channel measurements by $\{\mathbf{q}_i, g_i=\widetilde{G}(\mathbf{q}_i)\}_{i=1}^{M}$, where $g_i$ is the channel measurement, and $\widetilde{G}(\mathbf{q}_i)$ is given by
	\begin{align}
		\widetilde{G}(\mathbf{q}_i) = G_{ \mathbf{q}_0 \rightarrow \mathbf {q}} + z(\mathbf{q}_i),
		\label{eq:3}
	\end{align}
	with $z(\mathbf{q}_i)$ being the measurement noise.

	With the channel measurements, Kriging interpolation~\cite{10430216} provides an efficient method for accurately obtaining the radio map. Since Kriging can produce the best geostatistical linear unbiased estimator~\cite{10416187}, the feasibility of Kriging-based radio map methods has been validated~\cite{maiti2023ordinary}. To be specific, the Kriging method estimates $\hat{G}(\mathbf{q})$ at an unknown location $\mathbf{q} \in \mathcal{D}$ by constructing
	\begin{align}
		\hat{G}(\mathbf{q})  = \sum_{i=1}^{M} g_i\lambda_i + \lambda_0,
		\label{eq:4}
	\end{align}
	where $\{\lambda_i\}_{i=1}^{M}$ are the Kriging weights derived from $M$ measurements to minimize the mean squared prediction error $\mathbb{E}[(\hat{G}(\mathbf{q}) - {G}_{\mathbf{q}_0 \rightarrow  \mathbf{q}} )^2]$.

	The above radio map construction is time-invariant, but the actual radio propagation environment is not. Any changes in environmental factors can lead to dynamic radio maps.
	Let $\{\mathbf{q}_i, g_i^{(t)}\}_{i=1}^{M}$ denote the set of channel measurements at time $t$.
	Based on historical measurements, the dynamic radio maps can be modeled as a dynamical system:
	\begin{align}
		\hat{G}^{(t+1)}(\mathbf{q}) = F\left( \hat{G}^{(t)}(\mathbf{q})\right),
		\label{eq:5}
	\end{align}
	where $\{\hat{G}^{(t)}(\mathbf{q})\}_{\mathbf{q} \in \mathcal{D}}$ denotes the radio map at time $t$, while the evolution function $F(\cdot)$ is the focus of this study.
	The aim is to understand the dynamics of the radio map over time.
	
	In this paper, we consider a dynamic radio map scenario, where the dominant environmental factor is the moving transmitter.
	The channel dataset emulating mobility is generated by Sionna RT~\cite{sionna}, where the 3D scenario is provided by Sionna open-source library\footnote{https://nvlabs.github.io/sionna/api/rt.html\#example-scenes}.
	%Fig.~\ref{fig:3} shows the dynamic variations of the channel gain map over time.

	\section{Dynamic Channel Twinning Framework}\label{sec:3}
	In this section, we first provide an overview of the basis of DMD and the involved variants for our method. Then, the proposed Ens-DMD algorithm by combining cDMD and eDMD is presented. Finally, we summarize the dynamic radio map twinning framework.

	\subsection{Dynamic Mode Decomposition Basics}\label{sec:3.1}
	DMD is a data-driven technology that can solve complex spatio-temporal coupling systems, which can transform high-dimensional states into an inexpensive low-dimensional representation meanwhile allowing for insightful forecasting of dynamic behaviors. These merits enable DMD to analyze system dynamics in various applications.

	Assume $N$ measurements or snapshots collected from a dynamical system, where each data snapshot should be reshaped into a column vector, e.g., $\mathbf{g}^{(t)} = \left[g_1^{(t)}, g_2^{(t)}, \cdots, g_{M}^{(t)}\right]^{\mathsf{T}}\in \mathbb{R}^{M}$. The dimension of $M$ is the number of spatial points, which is equivalent to the number of positions within the radio map discussed in this paper.
	Accordingly, the data can be grouped into the following two matrices:
	\begin{align}
		\mathbf{G}_{1}^{N-1} &= \left[\mathbf{g}^{(1)}, \mathbf{g}^{(2)}, \cdots, \mathbf{g}^{(N-1)}\right]; \label{eq:6} \\
		\mathbf{G}_{2}^{N} &= \left[\mathbf{g}^{(2)}, \mathbf{g}^{(3)}, \cdots, \mathbf{g}^{(N)}\right].
		\label{eq:7}
	\end{align}
	Given the measured data in~\eqref{eq:6} and \eqref{eq:7}, the DMD principle is to extract a set of dynamic modes by performing the spatio-temporal decomposition on the data for a given system.
	Specifically, the DMD principle assumes a linear map of the dynamic process in~\eqref{eq:5}. It employs the \emph{Koopman operator}, $\mathbf{A}_{\mathbf{g}}\in\mathbb{R}^{M \times M}$, to map the original nonlinear dynamics into an infinite-dimensional linear system~\cite{nathan2015multi}. The Koopman operator satisfies $\mathbf{g}^{(t+1)}\approx \mathbf{A}_{\mathbf{g}}\mathbf{g}^{(t)}$, and thus
	\begin{align}
		\mathbf{G}_{2}^{N}\approx \mathbf{A}_{\mathbf{g}}\mathbf{G}_{1}^{N-1}. \label{eq:8}
	\end{align}
	
	Directly computing $\mathbf{A}_{\mathbf{g}}=\mathbf{G}_{2}^{N}\left(\mathbf{G}_{1}^{N-1}\right)^{\dagger}$ is tractable due to the large $M$ in practice. Instead, DMD calculates the proper orthogonal decomposition
	(POD) projection of $\mathbf{A}_{\mathbf{g}}$, thus resulting in a low-dimensional representation~\cite{10453222}:
	\begin{align}
		\widetilde{\mathbf{A}}_{\mathbf{g}}=
		\mathbf{U}_{\mathbf{g}}^{\mathsf H}
		\mathbf{G}_{2}^{N} \mathbf{V}_{\mathbf{g}} \mathbf{\Sigma}_{\mathbf{g}}^{-1}, \label{eq:9}
	\end{align}
	where $\mathbf{G}_{1}^{N-1}=\mathbf{U}_{\mathbf{g}} \mathbf{\Sigma}_{\mathbf{g}} \mathbf{V}_{\mathbf{g}}^{\mathsf H}$ based on singular value decomposition (SVD). Note that $\mathbf{U}_{\mathbf{g}} \in \mathbb{R}^{M\times r}$, $\mathbf{V}_{\mathbf{g}} \in \mathbb{R}^{(N-1)\times r}$, $\mathbf{\Sigma}_{\mathbf{g}} \in \mathbb{R}^{r\times r}$, and hence $\widetilde{\mathbf{A}}_{\mathbf{g}} \in \mathbb{R}^{r\times r}$. Generally, $r\ll M$ means a computationally efficient mode extraction.

	Applying the eigenvalue decomposition (EVD) of $\widetilde{\mathbf{A}}_{\mathbf{g}}$ yields $\widetilde{\mathbf{A}}_{\mathbf{g}} \mathbf{W}_{\mathbf{g}}=\mathbf{W}_{\mathbf{g}}\mathbf{\Lambda}_{\mathbf{g}}$, where $\mathbf{W}_{\mathbf{g}}$ and $\mathbf{\Lambda}_{\mathbf{g}}$ denote the eigenvector and eigenvalue matrices.
	Based on $\{\mathbf{U}_{\mathbf{g}}, \mathbf{W}_{\mathbf{g}}\}$, we can avoid explicitly computing $\mathbf{A}_{\mathbf{g}}$ to obtain its eigenvectors $\{\boldsymbol{\phi}_j\}_{j=1}^{r}$ and eigenvalues $\{\lambda_j\}_{j=1}^{r}$.
	
	As a result, DMD yields a system evolution process:
	\begin{align}
		\mathbf{g}^{(t+1)} = \sum_{j=1}^{r} \boldsymbol{\phi}_j \lambda_j^{t} b_j, \label{eq:10}
	\end{align}
	where $\{b_j\}_{j=1}^{r}$ denotes the initial amplitude of each mode and can be obtained from $\mathbf{g}^{(1)}$. Notably, $\{\boldsymbol{\phi}_j\}_{j=1}^{r}$, $\{b_j\}_{j=1}^{r}$, and $\{\lambda_j^{t}\}_{j=1}^{r}$ represent the DMD \emph{modes}, \emph{amplitudes}, and underlying \emph{evolution} rule (or dynamics), respectively. 
	DMD has an alternative equivalent form to \eqref{eq:10}, as follows:
	\begin{align}
		\mathbf{G}_{1}^{N-1} = \mathbf{\Phi}_{\mathbf{g}} \mathbf{B}_{\mathbf{g}} \mathbf{\Upsilon}_{\mathbf{g}}, 
		\label{eq:11}
	\end{align}
	where $\mathbf{\Phi}_{\mathbf{g}}=\left(\bm{\phi}_1, \bm{\phi}_2, \cdots, \bm{\phi}_r\right) \in \mathbb{C}^{M\times r}$ consists of the DMD modes, the diagonal matrix $\mathbf{B}_{\mathbf{g}}=\mathrm{diag}\left([b_1, b_2,\cdots,b_r]\right) \in \mathbb{C}^{r\times r}$ contains the amplitudes associated with each DMD mode, and the matrix $\mathbf{\Upsilon}_{\mathbf{g}} \in \mathbb{C}^{r\times (N-1)}$ captures the temporal evolution behavior of the DMD modes. The Vandermonde matrix $\mathbf{\Upsilon}_{\mathbf{g}}$ is structured as follows:
	\begin{align}
		\mathbf{\Upsilon}_{\mathbf{g}}=
		\begin{pmatrix}
			1 & \lambda_1 &\cdots&\lambda_1^{N-1} \\
			1 & \lambda_2 &\cdots&\lambda_2^{N-1} \\ 
			\vdots & \vdots & \ddots &\vdots \\ 
			1 & \lambda_{r} &\cdots&\lambda_{r}^{N-1}
		\end{pmatrix}.
		\label{eq:12}
	\end{align}
	
	Provided that the initial state amplitude, along with the eigenvalues and eigenvectors of the Koopman operator, are known, it is theoretically possible to reproduce or predict the system state at any arbitrary time.

	\subsection{Proposed Ensemble DMD Algorithm}\label{sec:3.2}
	We introduce two DMD variants related to our proposed algorithm.
	
	\textbf{1) cDMD}: To efficiently apply DMD to high-dimensional measurement data, cDMD can reduce computational complexity while retaining the essential features of the system dynamics by employing compression techniques~\cite[Chapter~9]{kutz2016dynamic}. Assuming full-state measurement $\mathbf{g}^{(n)}, \forall n=1,2,\cdots, N$ is available, cDMD captures the main dynamics of the datasets $\mathbf{G}_{1}^{N-1}$ and $\mathbf{G}_{2}^{N}$ through low-rank approximation. Specifically, we compress $\mathbf{G}_{1}^{N-1}$ and $\mathbf{G}_{2}^{N}$ to $\mathbf{X}_{c}$ and $\mathbf{X}'_{c}$ by
	\begin{align}
		\mathbf{X}_{c} &= \mathbf{C} \mathbf{G}_{1}^{N-1} \in \mathbb{R}^{p\times (N-1)}, \label{eq:13}\\
		\mathbf{X}'_{c} &= \mathbf{C} \mathbf{G}_{2}^{N}  \in \mathbb{R}^{p\times (N-1)}, \label{eq:14}
	\end{align}
	where $\mathbf{C} \in \mathbb{R}^{p\times M}$ is the carefully selected compression matrix with $r \le p\ll M$. The discussion on the construction of compression matrices can be found in~\cite{erichson2019compressed}. Apply standard DMD mentioned in Section~\ref{sec:3.1} to $\mathbf{X}_{c}$ and $\mathbf{X}'_{c}$, we can obtain the compressed DMD modes $\mathbf{\Phi}_{\mathbf{X}}$, and then reconstruct the DMD modes $\mathbf{\Phi}_{\mathbf{g}}$ by $\mathbf{\Phi}_{\mathbf{X}}=\mathbf{C}\mathbf{\Phi}_{\mathbf{g}}$.

	\textbf{2) eDMD}: Compared to traditional DMD, which operates only in the original state space, eDMD aims to construct a Koopman operator representation in a high-dimensional feature space~\cite[Ch.~10]{kutz2016dynamic}. The eDMD relies on basis functions to map the original state space to a high-dimensional feature space, extracting nonlinear behaviors and richer dynamics. 
	
	To avoid the curse of dimensionality, the kernel-based eDMD implicitly and efficiently constructs high-dimensional feature spaces. For two groups of data, $\{\mathbf{G}_{1}^{N-1}, \mathbf{G}_{2}^{N}\}$, the corresponding non-linear mapping of the data is given by
	\begin{align}
		\mathbf{\Psi}_{1} &= \left[ \bm{\psi}(\mathbf{g}^{1}), \bm{\psi}(\mathbf{g}^{2}) \cdots, \bm{\psi}(\mathbf{g}^{N-1})\right]; \label{eq:15}\\
		\mathbf{\Psi}_{2} &= \left[ \bm{\psi}(\mathbf{g}^{2}), \bm{\psi}(\mathbf{g}^{3}) \cdots, \bm{\psi}(\mathbf{g}^{N})\right], \label{eq:16}
	\end{align}
	where $\bm{\psi}(\mathbf{x})=\left[\psi_1(\mathbf{x}), \psi_2(\mathbf{x}), \cdots, \psi_J(\mathbf{x})\right]^{\mathsf{T}}$ collects a set of scalar basis functions with $J\gg M$. Similar to standard DMD, the Koopman operator $\mathbf{A}_{\text{eDMD}}$ can be obtained by solving the following problem~\cite{9147729}:
	\begin{align}
		\mathbf{A}_{\text{eDMD}} = \text{arg min}_{\mathbf{A}}\|\mathbf{\Psi}_2 - \mathbf{A} \mathbf{\Psi}_1 \|_F.
		\label{eq:17}
	\end{align}
	
\begin{figure}[t]
		\centering{}\includegraphics[width=3.5in]{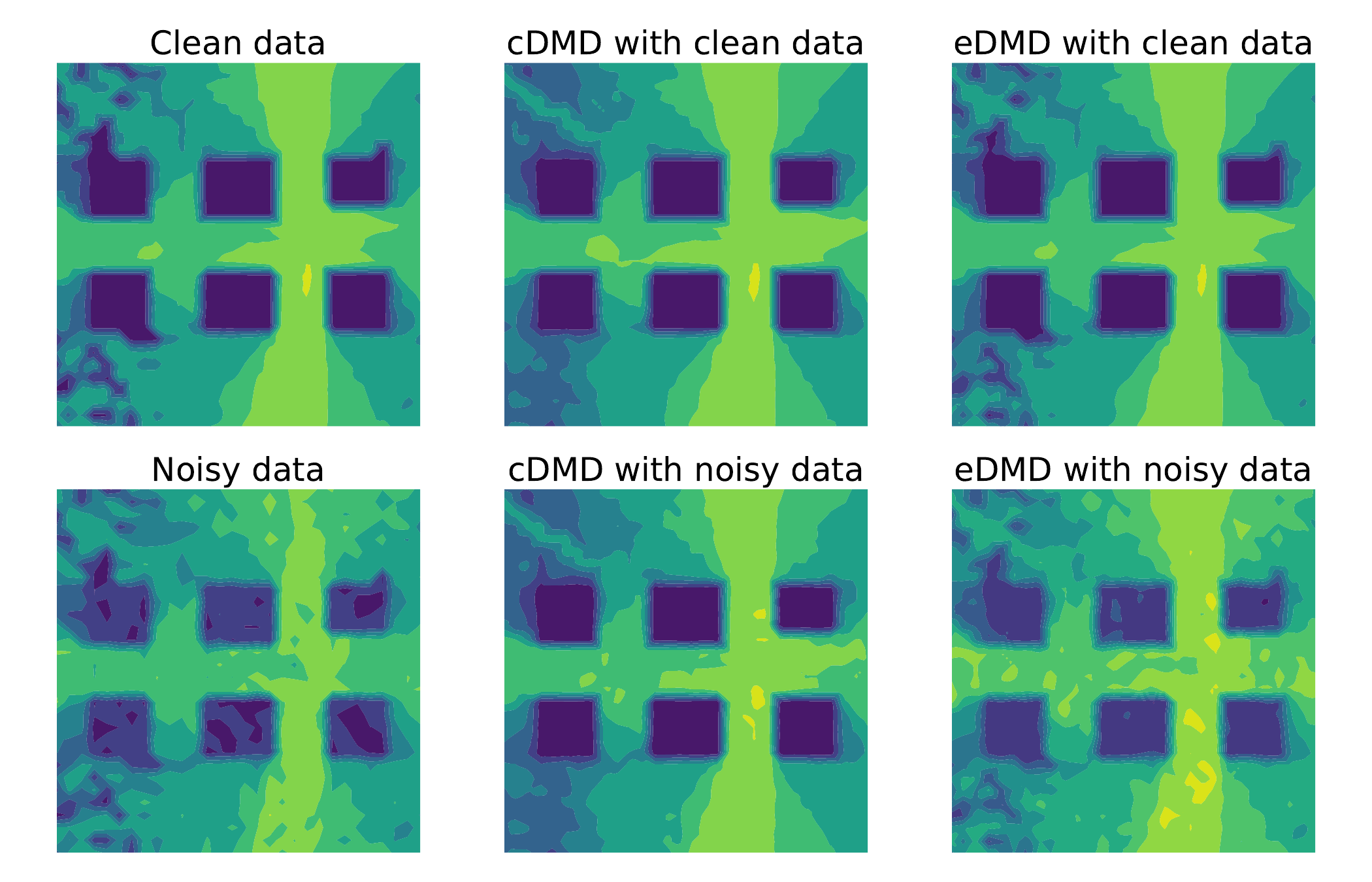}
		\caption{Performance comparison between cDMD and eDMD.}
		\label{fig:4}
	\end{figure}

	Since eDMD can capture nonlinear behaviors and complex dynamics in the high-dimensional feature space, it is capable of extracting fine-grained information from measurements compared to cDMD or standard DMD. For this reason, eDMD is well-suited for addressing noise-free radio map twinning tasks. The twin radio map results by using clean measurements are shown in Fig.~\ref{fig:4}(a). However, the twin performance of the eDMD is degraded with the noisy measurements, because eDMD treats noise behaviors as transient features to be captured. In this regard, cDMD exhibits relatively low sensitivity to noise behaviors, due to its insufficient ability to capture transient features. From Fig.~\ref{fig:4}(b), the twin radio map by eDMD is heavily affected by noise.

	\textbf{3) Proposed Ens-DMD: }	
	In light of the observations above, we propose an Ens-DMD algorithm, which effectively integrates the benefits of the two aforementioned DMD methods. Specifically, we first apply cDMD and eDMD on the snapshots to obtain their respective modes, amplitudes, and evolution rules, as given by
	\begin{align}
		\mathbf{G}_{1}^{N-1} &= \mathbf{\Phi}_{\mathbf{g}, \text{C}} \mathbf{B}_{\mathbf{g}, \text{C}} \mathbf{\Upsilon}_{\mathbf{g}, \text{C}}, \label{eq:18} \\
		\mathbf{G}_{1}^{N-1} &= \mathbf{\Phi}_{\mathbf{g}, \text{E}} \mathbf{B}_{\mathbf{g}, \text{E}} \mathbf{\Upsilon}_{\mathbf{g}, \text{E}}, \label{eq:19}
	\end{align}
	where $\mathbf{\Phi}_{\mathbf{g}, \text{C}}$ ($\mathbf{\Phi}_{\mathbf{g}, \text{E}}$), $\mathbf{B}_{\mathbf{g}, \text{C}}$ ($\mathbf{B}_{\mathbf{g}, \text{E}}$), and $\mathbf{\Upsilon}_{\mathbf{g}, \text{C}}$ ($\mathbf{\Upsilon}_{\mathbf{g}, \text{E}}$) represent the cDMD (eDMD) modes, amplitudes, and temporal evolution, respectively.

	\begin{figure*}[t]
		\centering{}\includegraphics[width=6.0in]{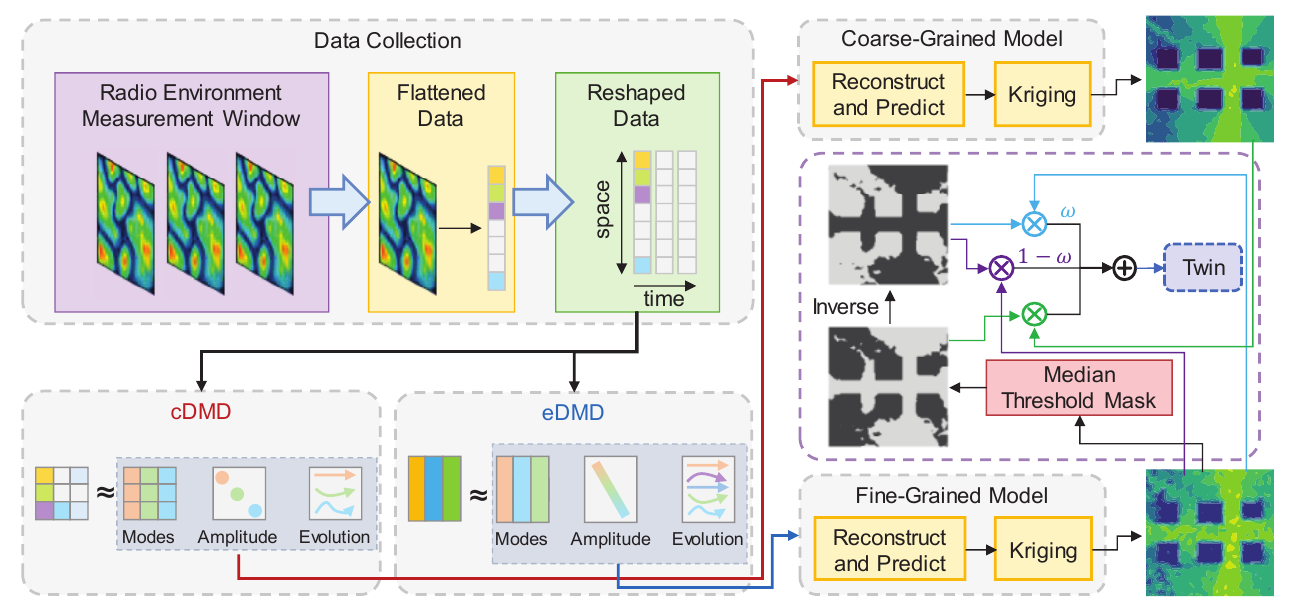}
		\caption{Dynamic radio map twinning framework. The framework consists of four key phases: (1) data collection from time series measurements, organized into a spatio-temporal matrix; (2) DMD application, utilizing both coarse-grained (cDMD) and fine-grained (eDMD) models to extract mode evolution; (3) spatial interpolation using Kriging to generate high-resolution radio maps; and (4) median threshold-based masking, which fuses stable and transient features from interpolated radio maps of two models to construct the final dynamic radio map.}
		\label{fig:5}
	\end{figure*}

	Leveraging the DMD modes, amplitudes, and evolution, we can further integrate Kriging interpolation to facilitate the reconstruction or prediction of the radio map state. As shown in Fig.~\ref{fig:5}, the twin maps by cDMD and eDMD after Kriging interpolation are denoted by $\hat{\mathbf{G}}_{\text{C}}$ and $\hat{\mathbf{G}}_{\text{E}}$, respectively. The twin map of $\hat{\mathbf{G}}$ is structured by
	\begin{align}
		\hat{\mathbf{G}}  = 
		\begin{pmatrix}
			\hat{G}(\mathbf{q}_{1,1})& \hat{G}(\mathbf{q}_{1,2})&{\cdots}&\hat{G}(\mathbf{q}_{1,Q_{y}})\\
			\hat{G}(\mathbf{q}_{2,1})& \hat{G}(\mathbf{q}_{2,2})&{\cdots}&\hat{G}(\mathbf{q}_{2,Q_{y}})\\
			{\vdots}&{\vdots}&{\ddots}&{\vdots}\\
			\hat{G}(\mathbf{q}_{Q_{x},1})& \hat{G}(\mathbf{q}_{Q_{x},2})&{\cdots}&\hat{G}(\mathbf{q}_{Q_{x},Q_{y}})
		\end{pmatrix}, \label{eq:20}
	\end{align}
	where $\hat{G}(\mathbf{q}_{i,j}), \forall i=1,2,\cdots,Q_x; j=1,2,\cdots, Q_y$ stems from the Kriging interpolation~\eqref{eq:4}, and $Q_x \times Q_y$ represents the dimensions of the output grid of Kriging.

	Note that cDMD captures the stable variations of the system, thus ignoring transient noise features. Conversely, eDMD often produces subtle features and transient fluctuations, which inevitably involve high-frequency noise. 
	To filter the noise from the eDMD map $\hat{\mathbf{G}}_{\text{E}}$, we design a median-threshold mask method. The idea behind the median-threshold mask method is to preserve the large-timescale features from noise-insensitive cDMD, while absorbing some small-timescale features from eDMD. The median of the eDMD map $\hat{\mathbf{G}}_{\text{E}}$ is denoted by $m_{G} = \mathrm{Median}(\hat{\mathbf{G}}_{\text{E}})$.
	Then, the median-threshold mask for eDMD, denoted by $\mathbf{M}_{G}$, is defined as
	\begin{align}
		M_{G}(k,l)  = \begin{cases}
			1, & \hat{\mathbf{G}}_{\text{E}}(k,l) \ge m_{G}; \\
			0, & \hat{\mathbf{G}}_{\text{E}}(k,l)  <  m_{G}.
		\end{cases} \label{eq:21}
	\end{align}
	Thus, the corresponding inverse mask can be readily obtained by $\bar{\mathbf{M}}_{G}=1-\mathbf{M}_{G}$.
	
	To preserve the large-timescale stable features from cDMD, we utilize the median-threshold mask to filter the cDMD map $\hat{\mathbf{G}}_{\text{C}}$. The resulting large-timescale feature map is given by
	\begin{align}
		\hat{\mathbf{G}}_{0} = \hat{\mathbf{G}}_{\text{C}} \odot \mathbf{M}_{G}. \label{eq:22}
	\end{align}	
	Consider the weighted fusion of small-timescale features from two DMD methods to achieve the twin representation of local transient features for noise suppression. 
	The small-timescale feature maps of cDMD and eDMD are given by
	\begin{align}
		\hat{\mathbf{G}}_{1,\text{C}} = \hat{\mathbf{G}}_{\text{C}} \odot \bar{\mathbf{M}}_{G}; \label{eq:23} \\
		\hat{\mathbf{G}}_{1,\text{E}} = \hat{\mathbf{G}}_{\text{E}} \odot \bar{\mathbf{M}}_{G}. \label{eq:24}
	\end{align}
	
	At the end, the twin radio map is generated using the following equation:
	\begin{align}
		\mathbf{G}_{\text{twin}} = \hat{\mathbf{G}}_{0} + \omega \hat{\mathbf{G}}_{1,\text{C}} + (1-\omega) \hat{\mathbf{G}}_{1,\text{E}}, \label{eq:25}
	\end{align}
	where $\omega$ is the weight coefficient that determines the relative contributions of the two DMD outputs to the final radio map. From~\eqref{eq:25}, the proposed method can be regarded as an ensemble of two DMD distinct algorithms.
	Above all, we analyze the advantages of two DMD algorithms and fuse their features to enhance the overall performance. Hence, we name this approach the ensemble DMD algorithm. By leveraging the complementary traits of both algorithms, the Ens-DMD algorithm achieves enhanced accuracy and robustness in the presence of non-negligible noise.

	\subsection{Overall Dynamic Radio Map Twinning Framework}\label{sec:3.3}
	% The detailed algorithmic description involved in the developed dynamic radio map twinning framework has been presented in the previous subsection. 
	In this section, we summarize the overall dynamic radio map twinning framework, as illustrated in Fig.~\ref{fig:5}.	
	The framework encompasses the following key phases: data collection, DMD, spatial interpolation, and median-based threshold masking.
	\subsubsection{Data collection} In the considered time window length, the channel gain measurements at different time instances are collected from spatially distributed meshes. Then, each measurement data needs to be reshaped into a column vector $\{\mathbf{g}^{(t)}\}_{t=1}^{N}$. The reshaped data is organized into a spatio-temporal data matrix frame (i.e., $\mathbf{G}_{1}^{N-1}$ and $\mathbf{G}_{2}^{N}$), as required by DMD.
	
	\subsubsection{DMD} After the data preparation phase, we independently apply the cDMD and eDMD algorithms to the collected data matrix. This allows us to extract and store the modes, amplitudes, and evolution rules for each algorithm at the BS server. 
	
	\subsubsection{Spatial interpolation} As discussed in Section~\ref{sec:3.2}, the eDMD is responsible for capturing fine-grained features, whereas the cDMD focuses on the coarse-grained features. The resulting DMD decomposition at various levels (timescales) is subsequently employed in Kriging interpolation to create both coarse-grained and fine-grained radio map models.
	The fine-grained decomposition results enhance the spatial resolution of the radio map, while the coarse-grained representation serves to filter out the measurement noise.
	
	\subsubsection{Median threshold-based masking} To leverage the advantages of both DMD models, we implement the Ens-DMD algorithm. The median of the radio map attained by eDMD is first derived, and thus median-based threshold mask $\mathbf{M}_G$ can be constructed. This mask is then applied to the cDMD-based radio map to extract stable features. Next, the inverse mask $\bar{\mathbf{M}}_G$, which is complementary to $\mathbf{M}_G$,  extracts the transient features from the eDMD-based radio map. To mitigate the noise impact associated with the fine-grained model,  we also apply $\bar{\mathbf{M}}_G$ on the cDMD-based radio map.
	This results in a weighted fusion of the extracted features from both models, as described in Eqn.~\eqref{eq:25}.

	Through these phases, the proposed dynamic radio map twinning framework generates the radio map for the specified time instance. As the operational time increases, the measurement window can incorporate a new set of observation data, enabling the DMD to refine its learning regarding the dynamic system.

    \begin{remark}
    DMD and the encoder-decoder (E2D) framework share a fundamental workflow by projecting data sequences into a compressed representation (encoding) and subsequently reconstructing them (decoding)~\cite{naiman2023operator}. While recent findings demonstrate that E2D architectures in large language models (LLMs) conform to Koopman operator principles~\cite{mezic2023operatormodel}, DMD fundamentally differs in its mathematical foundation and interpretability compared to E2D.
    According to Koopman theory, DMD extracts physically interpretable coherent spatial modes during the compression step. Singular value decomposition or dominant mode projection is utilized to \emph{encode} the nonlinear system states into linearizable observables in an alternative domain. In the reconstruction step, these modes and their temporal evolution are employed to \emph{decode} the full state dynamics, enabling high-fidelity transformation from the encoded domain back to the original input space~\cite{pmlrdey23a}.
    In contrast to neural network-based E2D models, the encoding step in DMD explicitly decomposes the system into stable and unstable subspaces via spectral analysis, enabling targeted analysis of transient growth or decay. The decoding step then linearly recombines these components to restore the complete state, thereby avoiding black-box transformations~\cite{okorokov2024explaining}. This inherent E2D structure emerges naturally from the mathematical formulation of DMD, predating modern E2D paradigms.

    These structural similarities also open promising avenues for hybrid paradigms. For instance, generative artificial intelligence (GAI)~\cite{10757328, 10841365} can utilize multimodal prompts (e.g., text and images) to synthesize high-resolution radio maps from sparse measurements, addressing the data quality constraints of DMD methods. Moreover, leveraging radio propagation models as regularization terms or feature extractors can enhance the robustness of DMD-based channel twinning in complex environments. For example, physics-regularized loss functions~\cite{10634040} effectively integrate model-based priors with data-driven techniques.
    \end{remark}
	
	\subsection{Benchmark Schemes and Complexity Analysis}\label{sec:3.4}
		Four representative methods serve as the benchmarks for the proposed Ens-DMD algorithm:
		\begin{itemize}
			\item DMD~\cite{10279471}: A standard DMD algorithm, combined with Kriging interpolation, is employed for radio map reconstruction and prediction. 
			%DMD captures the modes and evolution rules from the original data, while Kriging interpolation is used to construct the complete radio map.
			\item KKF~\cite{9931518}: As a prominent radio map estimation method, KKF integrates Kalman filtering with Kriging interpolation to reconstruct and predict spatial-temporal data. 
			%The approach enhances accuracy by addressing both spatial and temporal correlations in radio map predictions. 
			\item DMC~\cite{9741316}: A dynamic matrix completion (DMC) method with Kalman filter prediction is considered, where the radio map is recovered via nuclear norm minimization.
			\item LSTM~\cite{10292913}: A low-complexity LSTM network is utilized for radio map reconstruction. The network architecture consists of an encoder with an $L$-unit LSTM cell, which takes known $\{\mathbf{g}^{(t)}\}_{t=1}^{N}$ and the locations to be interpolated (set to 0) as input, followed by a decoder formed by a dense layer with rectified linear unit (ReLU) activation that predicts the radio map.
	\end{itemize}

		We compare the computational complexity of the proposed Ens-DMD algorithm to benchmark methods. Consider an $N \times M$ spatio-temporal data matrix, where $N$ is the state dimension per snapshot and $M$ is the number of snapshots. The resolution of the reconstructed map is denoted by $Q$.

		For the KKF algorithm, the Kalman filter processes each snapshot sequentially, yielding the total complexity across $M$ snapshots is $\mathcal{O}(MN^3)$~\cite{7194838}.
		For the DMD algorithm, the complexity primarily arises from the SVD operation, yielding a complexity of $\mathcal{O}(NM^2)$~\cite{10453222}. The overall complexity order of the LSTM method is given by $O(QL^2 +LQ^2)$~\cite{10292913}.

		As per interpolation complexity for each time step, both KKF and Ens-DMD algorithms incorporate the ordinary Kriging, which incurs an interpolation complexity of $\mathcal{O}(N^2Q^2)$~\cite{maiti2023ordinary}. The matrix completion in DMC involves nuclear norm minimization with an interpolation complexity of $\mathcal{O}(N^2Q^3)$~\cite{9741316,9992123}.

	%the Kalman filter operates as a recursive method for state estimation that requires matrix multiplication and inversion. At each time step, the Kalman filter updates both the system state and the corresponding covariance matrix. Given that each time step primarily involves the matrix multiplication and inversion of a square covariance matrix with dimensions $N \times N$, the computational complexity of this step is $\mathcal{O}(N^3)$~\cite{7194838}.
	
	%The computational complexity order of Ordinary Kriging is O(N3) [29], due to the inversion of a semivariance matrix to compute the weights.
	
	%the complexity of DMD depends mainly on the SVD of the data matrix ($M>N$) to extract the dynamic modes of the system. Therefore, 
	
	%For comparison, Table~\ref{tb:1} summarizes the computational complexity of the proposed method and existing methods.

	%\section{Comparative Metrics and Complexity of Radio Map Twinning}\label{sec:4}

	\section{Simulation Analysis}\label{sec:4}	
	In this section, we present numerical examples to assess the performance of the proposed twin algorithms for radio maps under noise-free and noisy dataset conditions. Additionally, we explore a power allocation application scenario to illustrate the significance of various twin performance metrics.

	The 3D map scene of a simple street is generated using the Sionna platform~\cite{sionna}. Consider a 150 m by 150 m coverage region, with grid points uniformly sampled at a spacing of 5 m, as illustrated in Fig~\ref{fig:2}(b). 
	The comprehensive wireless propagation phenomena incorporates multipath propagation (with a maximum depth of three bounces), diffraction, and scattering effects. These propagation characteristics are enabled in the NVIDIA Sionna platform.
	Notably, Sionna provides a cost-effective and controllable environment for simulating real-world channel behaviors~\cite{zubow2024ns3}.
    We collect multiple snapshots of path gain measurements by empirically simulating the mobility of the MD through the Sionna RT module~\cite{10465179}. By setting the carrier frequency to 28 GHz, we utilize Sionna to generate environment-aware true channel gain values for 500 snapshots, with the MD moving at a displacement speed of 50 m/s. By default, the time interval between snapshots is $\delta_t=2$ ms. Note that the transmitter is moving throughout the simulation, emulating realistic mobility scenarios.
    We set $M=30 \times 30$ and $N=20$ for subsequent algorithm implementation.
	% algorithm parameter settings
	For the noisy dataset, the measurement noise follows a Gaussian distribution with zero-mean and variance 10 throughout~\cite{5484600}.
	In the Kriging interpolation, the returned dimension is set to be $Q_x \times Q_y =100 \times 100$.
	In addition, we implement the DMD algorithms using the PyDMD package~\cite{demo18pydmd}.
	For the cDMD algorithm, we set the DMD rank $r=5$, which offers appealing benefits, e.g., enhanced computational efficiency, reduced storage requirements, and improved stability.

	\subsection{Comparative Metrics}\label{sec:4.1}
	To comprehensively evaluate the twin quality of radio maps, we define four metrics: mean squared error (MSE), peak signal-to-noise ratio (PSNR), structural similarity index (SSIM), and correlation.
	\begin{itemize}
		\item MSE is defined as the average of the squared differences between each pixel of the true map $\mathbf{G}_{\text{true}}$ and the twin map $\mathbf{G}_{\text{twin}}$:
		\begin{align}
			\text{MSE} = \frac{\sum_{k=1}^{Q_x} \sum_{l=1}^{Q_y} \left(\mathbf{G}_{\text{true}}(k,l) - \mathbf{G}_{\text{twin}}(k,l) \right)^2}{Q_x Q_y} . \label{eq:26}
		\end{align}	   
		
		\item PSNR is defined as a ratio derived from MSE, measured in dB, and is calculated as follows: 
		\begin{align}
			\text{PSNR} = 10 \log_{10} \left(\frac{\vert\max(\mathbf{G}_{\text{true}})\vert^2}{\text{MSE}}\right). \label{eq:27}
		\end{align}
		Similar to MSE, it concentrates solely on the errors between corresponding pixels.
		
		\item SSIM measures the similarity between two radio maps, as defined by
		\begin{align}
			\text{SSIM} = \frac{\left(2 \mu_{G} \mu_{\hat{G}} + c_1\right) \left( 2\sigma_{G\hat{G}}+c_2\right)}{\left(\mu_{G}^2 + \mu_{\hat{G}}^2 + c_1\right) \left(\sigma_{G}^2+\sigma_{\hat{G}}^2+c_2\right)}, \label{eq:28}
		\end{align}
		where $\mu_{G}$ and $\mu_{\hat{G}}$ are the means of true and twin maps, respectively; $\sigma_{G}$ and $\sigma_{\hat{G}}$ are the  variances of true and twin maps, respectively; $\sigma_{G\hat{G}}$ is the covariance between the true and twin maps; and $c_1$ and $c_2$ indicate the constants introduced for stability.  
		Since SSIM reflects the map structure information, it is a more effective metric than MSE in capturing the sensing quality of channel gain maps and especially environment sensing tasks~\cite{9729785}.
		\item Correlation measures the similarity between the pixel values of two maps, and is defined as
		\begin{align}
			\text{Corr} = \frac{ \mathbf{g}_{\text{true}}^{\mathsf{T}} \mathbf{g}_{\text{twin}} }{\Vert \mathbf{g}_{\text{true}} \Vert_{2}  \Vert \mathbf{g}_{\text{twin}}\Vert_{2}}, \label{eq:29}
		\end{align}
		where $\bar{\mathbf{G}}_{\text{true}}$ and $\bar{\mathbf{G}}_{\text{twin}}$ are the average values of true and twin maps, respectively; and $\mathbf{g}_{\text{true}} = \mathrm{vec}\left(\mathbf{G}_{\text{true}} - \bar{\mathbf{G}}_{\text{true}}\right)$, and $\mathbf{g}_{\text{twin}} =\mathrm{vec}\left(\mathbf{G}_{\text{twin}} - \bar{\mathbf{G}}_{\text{twin}}\right)$.
		Correlation focuses solely on the relationship between pixel values, and cannot reflect the overall structural information of maps.
	\end{itemize}
	It is noteworthy that quantifying the similarity of twin radio maps can be approached in two ways: By comparing the overall pixel-to-pixel error, or by analyzing the differences in channel gain distribution structures. The suitability of each performance metric is contingent upon the specific application requirements of wireless communications.
	
	%Algorithm run-time against the number of snapshots
	\begin{figure}[t]
		\centering{}\includegraphics[width=3.5in]{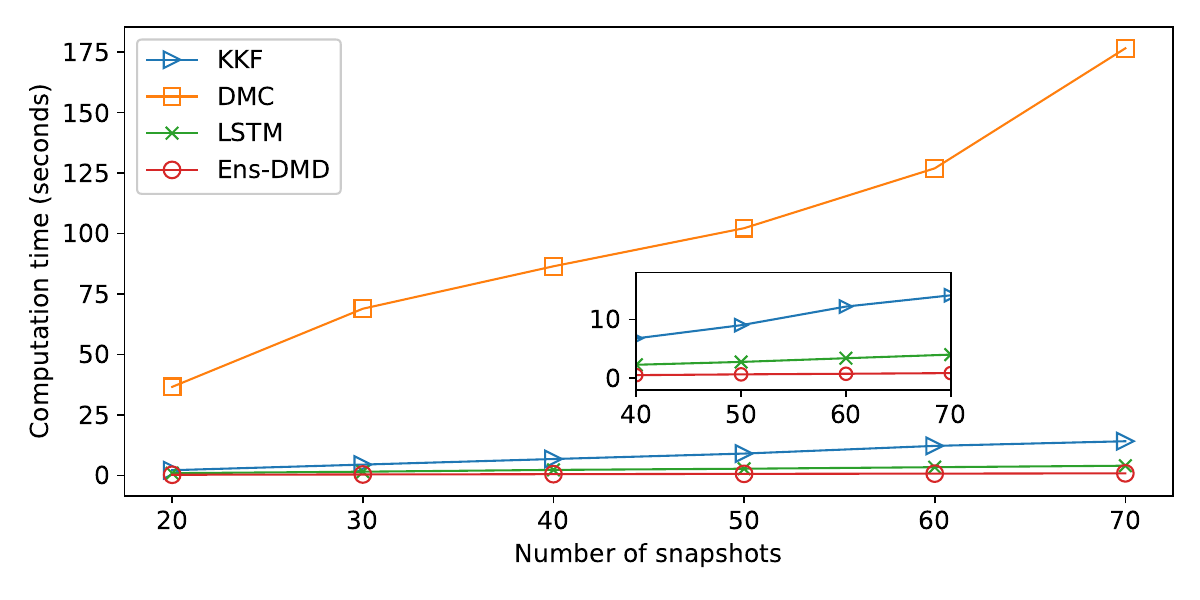}
		\caption{Algorithm run-time against the number of snapshots.}
		\label{fig:6}
	\end{figure}
	
	\begin{figure*}[t]
		\centering{}\includegraphics[width=6.0in]{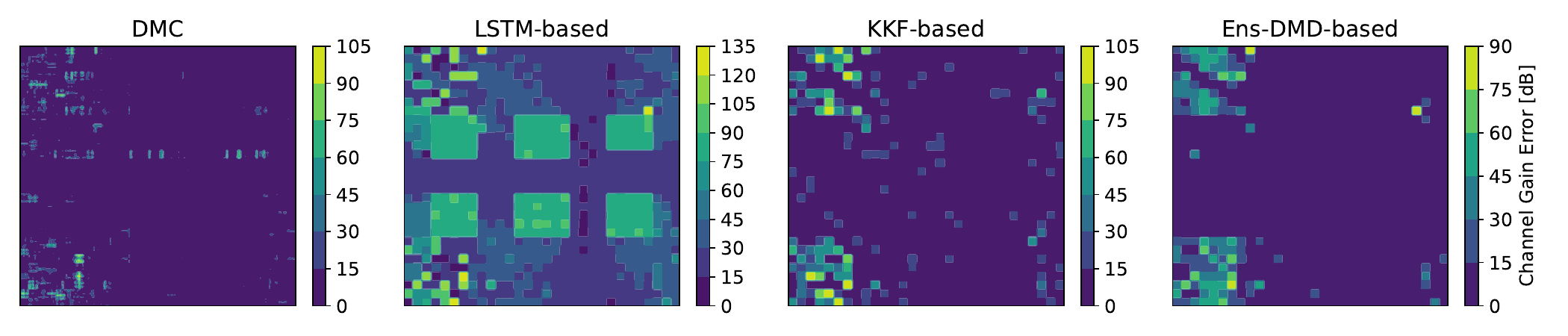}
		\caption{Absolute error maps for different radio map twinning methods.}
		\label{fig:7}
	\end{figure*}

	\subsection{Computational Efficiency and Parameter Sensitivity}\label{sec:4.2}
	Fig.~\ref{fig:6} explores the running time of different algorithms. We see that the Ens-DMD consumes the lowest running time, whereas DMC requires a significant amount of computational time.
	The running time of Ens-DMD increases very slowly with the growing number of snapshots. Notably, the Ens-DMD remains below 1 second even as the snapshot dimensions rise from 20 to 60. 
	In stark contrast, the DMC exceeds 25 seconds at a snapshot dimension of 20, rendering it impractical for real-world applications. Additionally, the LSTM shows a lower running time compared to KKF.
	%We also obverse that the KKF exceeds 2 seconds at a snapshot dimension of 20. This is because the Kalman filter gives a total complexity of $\mathcal{O}(MN^3)$, which is considerably higher than the complexity of the SVD operation in the DMD, $\mathcal{O}(NM^2)$.
	
	%补充权重影响图和dmd rank影响图
	\begin{figure}[t]
		\centering
		\subfigure[]
		{\includegraphics[width=3.5in]{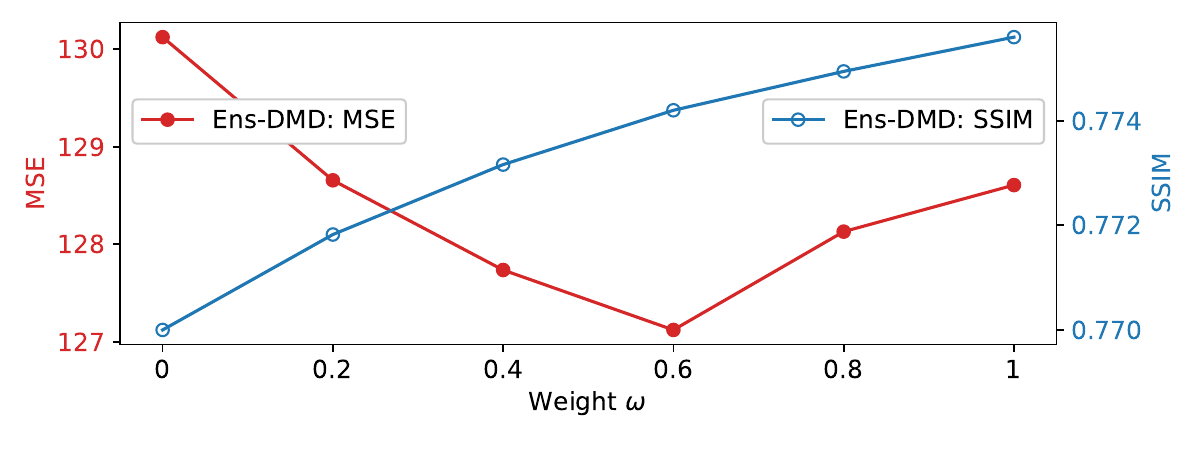}} 
		\subfigure[]
		{\includegraphics[width=3.5in]{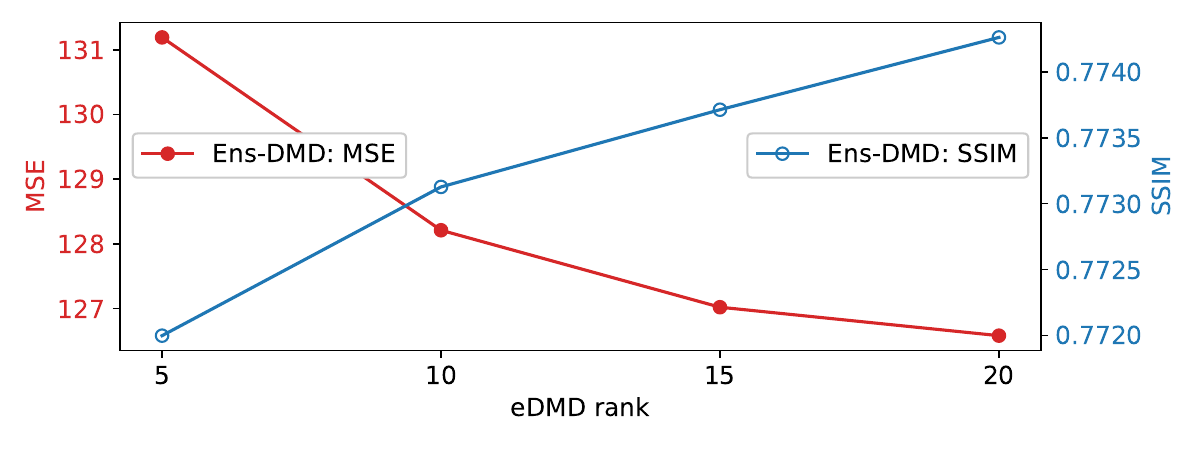}}
		\caption{Twin performance against (a) weight $\omega$ and (b) eDMD rank.}
		\label{fig:8}
	\end{figure}

	Fig.~\ref{fig:7} compares the map twinning error for different algorithms. While the DMC exhibits superior predictive performance with minimal occurrence of significant deviations, the high computational time prohibits its feasibility for time-varying scenarios. The accuracy of the LSTM is poor because the DL methods fail to effectively extract patterns from small-scale spatiotemporal datasets. By contrast, KKF and Ens-DMD offer impressive accuracy and desirable computational delays. In what follows, we compare KKF and Ens-DMD.

	Fig.~\ref{fig:8} examines the impacts of weight coefficient $\omega$ and eDMD rank on the algorithm performance. In Fig.~\ref{fig:8}(a), we observe that as the weight increases, the SSIM exhibits a monotonic increase, whereas the MSE initially decreases and subsequently rises. The inflection point occurs at $\omega=0.6$, indicating the optimal balance that simultaneously yields both minimal MSE and enhanced SSIM performance. This observation offers valuable insight for parameter selection: a suitable $\omega$ can be selected within the range of 0.4 to 0.8 to effectively balance MSE and SSIM. Fig.~\ref{fig:8}(b) illustrates that an increased DMD rank leads to improved twin performance. This is because eDMD approximates the infinite-dimensional Koopman operator by projecting the system dynamics onto a finite-dimensional space spanned by a selected dictionary of functions. Consequently, a higher DMD rank preserves more basis functions, thereby capturing a broader spectrum of dynamical features.

\begin{figure}[t]
		\centering{}\includegraphics[width=3.5in]{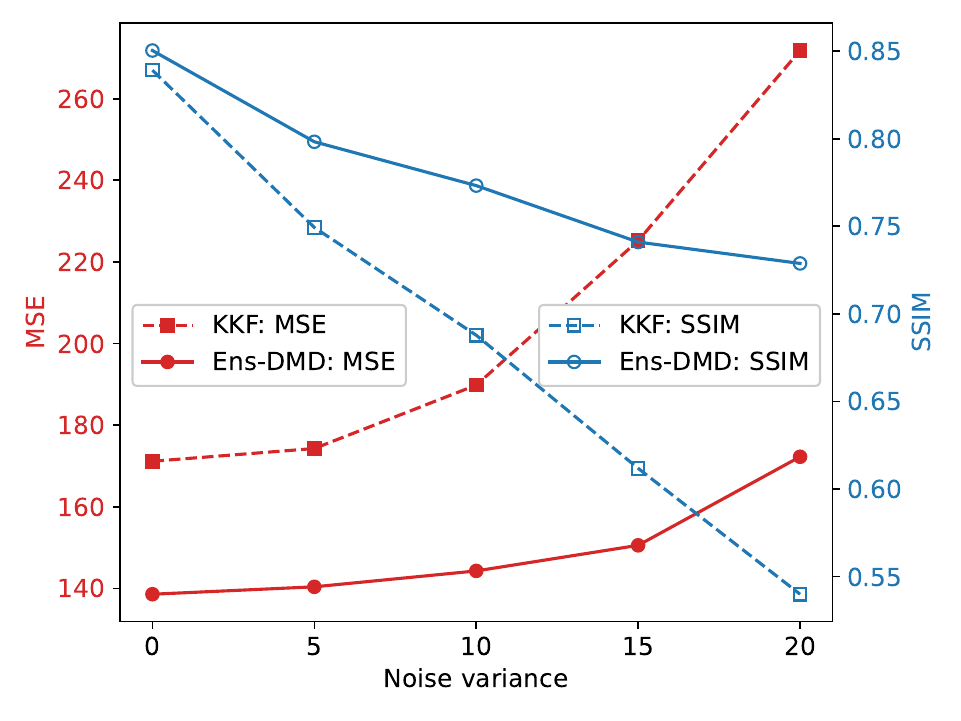}
		\caption{Prediction performance against noise variance.}
		\label{fig:9}
	\end{figure}

	To assess the impact of noise variance on the predictive performance of radio map twinning, Fig.~\ref{fig:9} compares the KKF and Ens-DMD algorithms using two metrics: MSE and SSIM. Both algorithms exhibit an upward trend in MSE with increasing noise variance, while the KKF algorithm demonstrates a more rapid increase. In terms of SSIM, both algorithms show a downward trend as noise variance increases. However, the decline in the Ens-DMD algorithm is considerably slower. Thus, the proposed Ens-DMD method provides significant advantages in pixel-level accuracy and the similarity of spatial distribution structures in channel gains.
	
	%% 不同噪声下
	
	\begin{figure}[t]
		\centering
		\subfigure[]
		{\includegraphics[width=3.5in]{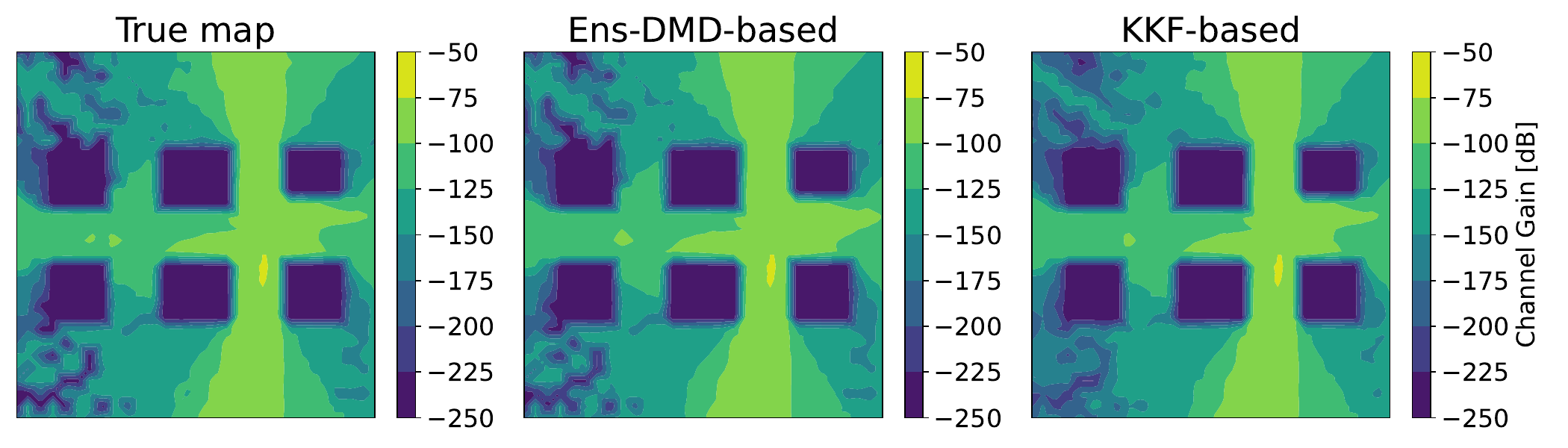}} 
		\subfigure[]
		{\includegraphics[width=3.5in]{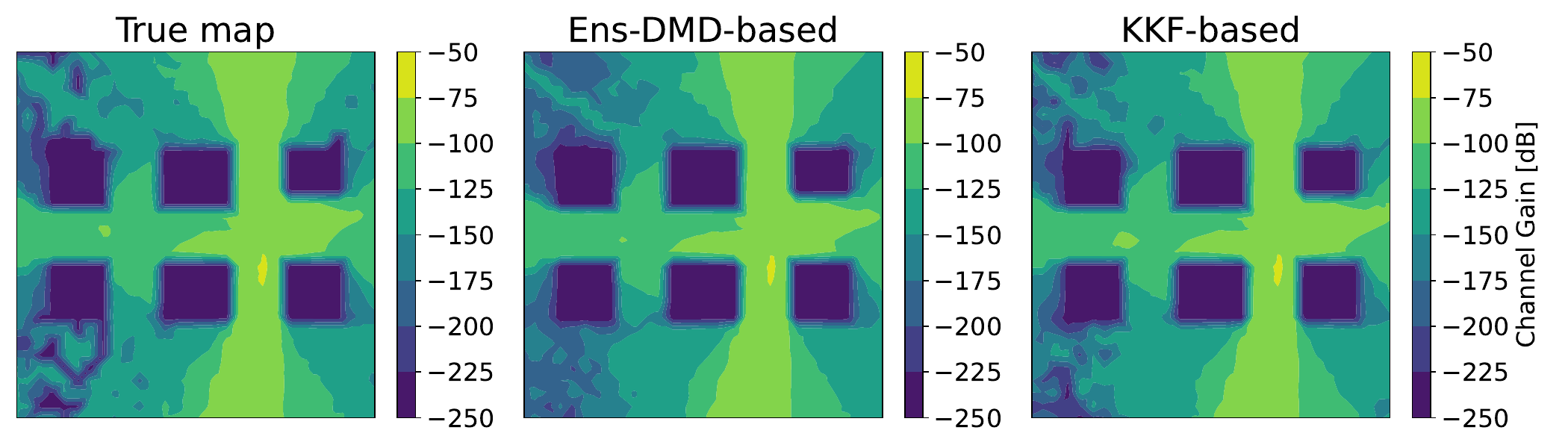}}
		\caption{Twin channel gain map with clean dataset: (a) Reconstruction, and (b) prediction.}
		\label{fig:10}
	\end{figure}

	\begin{figure}[t]
		\centering
		\subfigure[]
		{\includegraphics[width=2.5in]{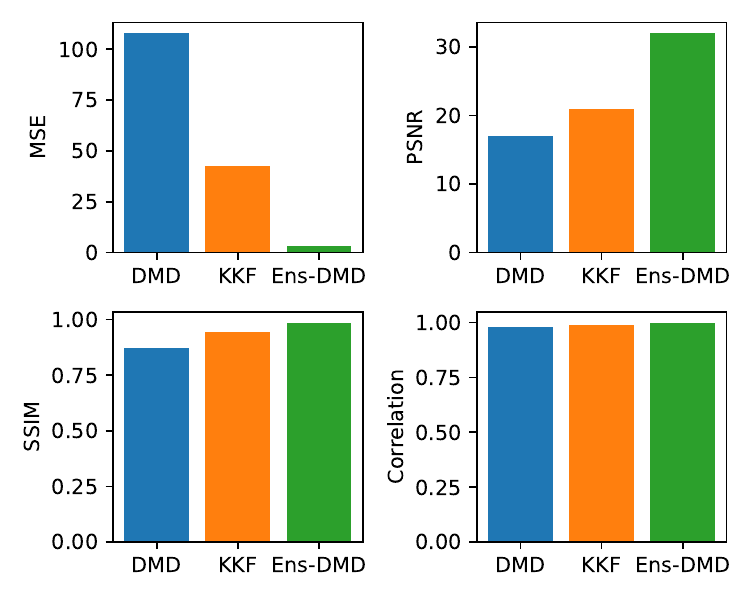}} 
		\subfigure[]
		{\includegraphics[width=2.5in]{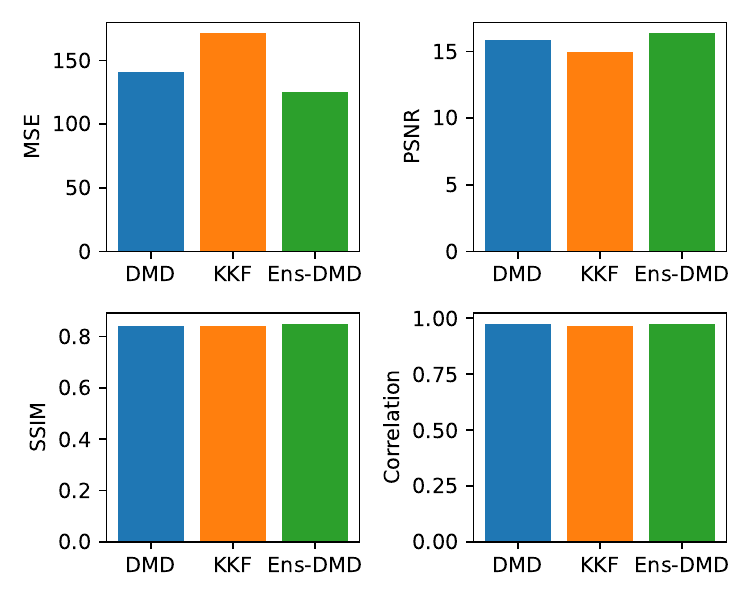}}
		\caption{Twin performance metric with clean dataset: (a) Reconstruction, and (b) prediction.}
		\label{fig:11}
	\end{figure}

	\subsection{Twin Performance under Different Scenarios}\label{sec:4.3}
	Fig.~\ref{fig:10} visualizes the twin channel gain maps based on 20 clean snapshots. Radio map twinning tasks include reconstruction of the original map state and prediction for future map state.
	Reconstruction holds potential meaningful applications in network monitoring, such as anomaly detection and fault diagnosis~\cite{10464447}.
	%% t=15
	Fig.~\ref{fig:10}(a) compares the reconstructed maps and ground truth, by taking $t=15 \delta_t$ as an example. It can be visualized that reconstructed radio maps produced by both Ens-DMD and KKF methods show good fidelity in channel gain distribution.
	%% t=25
	The visualization in Fig.~\ref{fig:10}(b) shows the prediction results for $t=25 \delta_t$. Since this is an unseen sample, the prediction performance is somewhat degraded compared to the reconstruction. The spatial distributions of channel gain for both algorithms exhibit a degree of deviation.

	With a clean dataset, Fig.~\ref{fig:11} presents a comparative analysis of the twin performance of standard DMD, KKF, and Ens-DMD algorithms across various metrics, which are defined in Section~\ref{sec:4.1}. It is observed in Fig.~\ref{fig:11}(a) that the proposed Ens-DMD algorithm outperforms the other two algorithms when evaluated on the selected reconstruction performance metrics. 
	The MSE, PSNR, and SSIM metrics clearly demonstrate a remarkable performance gap in evaluating the twin performance of radio maps. It is noteworthy that MSE is inversely related to PSNR, indicating that we can rely on either of these two metrics to assess the total reconstruction error. On the contrary, the correlation metric fails to effectively reflect the performance differences among the three algorithms, despite the presence of visually discernible differences.
	
	Fig.~\ref{fig:11}(b) illustrates the prediction performance results. 
	The Ens-DMD algorithm still demonstrates superior error performance compared to the other algorithms. Moreover, we observe that the standard DMD displays a lower overall error performance in comparison to the KKF. All three algorithms produce closely aligned SSIM values, indicating that the predicted radio maps possess similar structural characteristics.

\begin{figure}[t]
		\centering
		\subfigure[]
		{\includegraphics[width=3.5in]{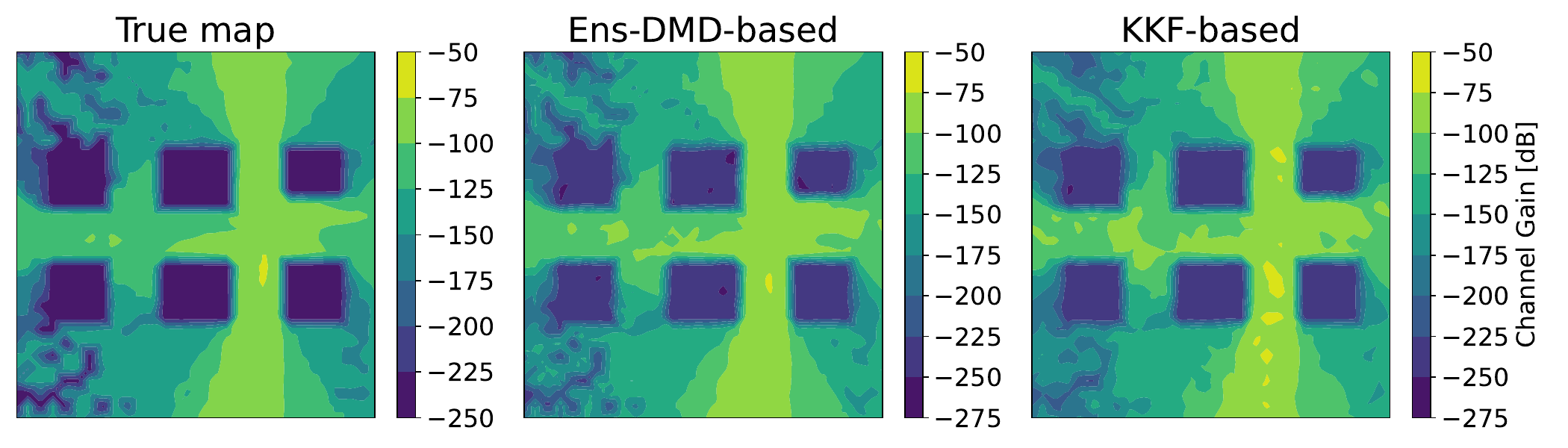}} 
		\subfigure[]
		{\includegraphics[width=3.5in]{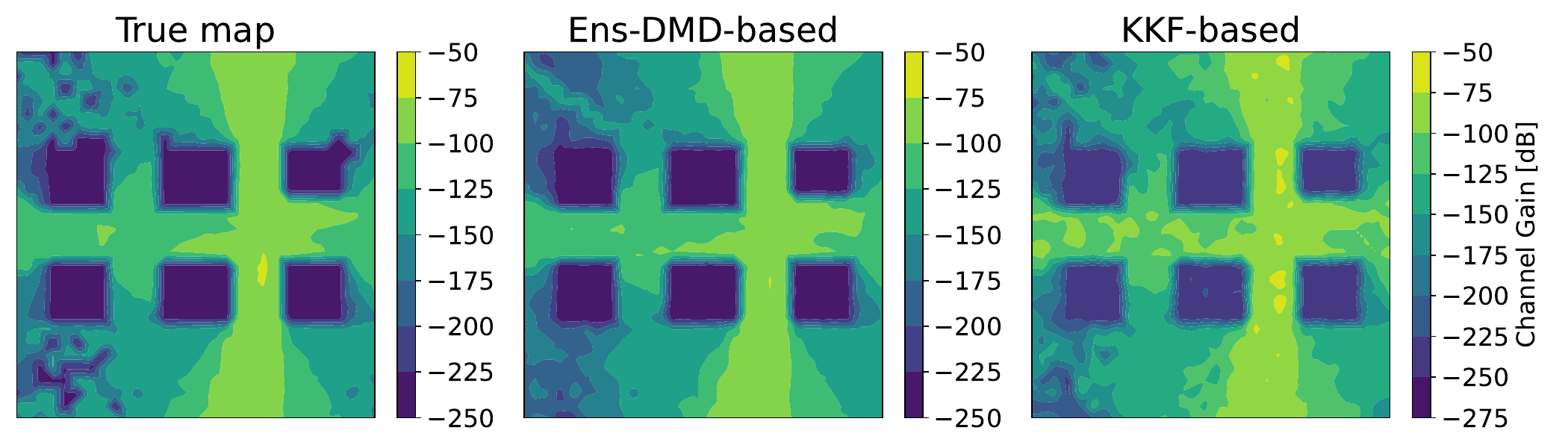}}
		\caption{Twin channel gain map with noisy dataset: (a) Reconstruction, and (b) prediction.}
		\label{fig:12}
	\end{figure}

	%\subsection{Twin Performance with Noisy Measurements}\label{sec:4.4}
	In Figs.~\ref{fig:12}(a) and~\ref{fig:12}(b), we visualize the reconstructed and predicted radio maps based on 20 noisy snapshots. From both reconstruction and prediction results, the radio map generated by KKF displays numerous noise artifacts, which disrupt the structural distribution of channel gain in the spatial domain. Conversely, the radio map generated by Ens-DMD demonstrates a strong resilience to noise.

	%%map上每一个点都可以用多普勒公式做一个推移，更新系数，得到基于模型的方法。

	%% 比各项metric
	\begin{figure}[t]
		\centering
		\subfigure[]
		{\includegraphics[width=2.5in]{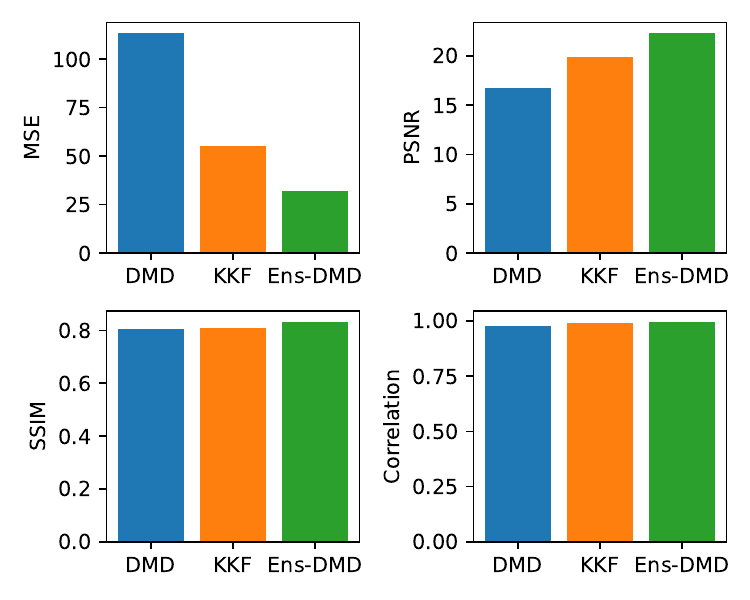}} 
		\subfigure[]
		{\includegraphics[width=2.5in]{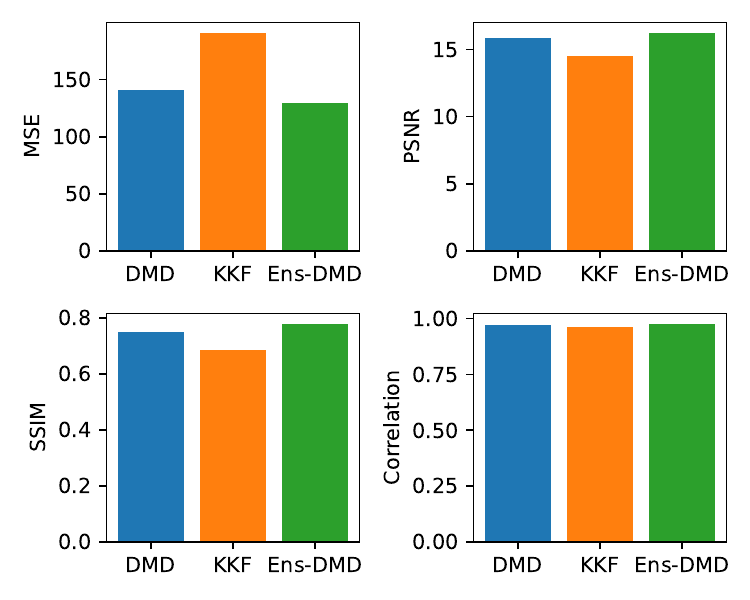}}
		\caption{Twin performance metric with noisy dataset: (a) Reconstruction, and (b) prediction.}
		\label{fig:13}
	\end{figure}

	On the noisy dataset, Fig.~\ref{fig:13} illustrates the comparative performance of the standard DMD, KKF, and Ens-DMD algorithms based on various metrics. Comparing Figs.~\ref{fig:11}(a) and~\ref{fig:13}(a), we see that the Ens-DMD algorithm consistently outperforms all other algorithms across all reconstruction metrics, regardless of whether the datasets are clean or noisy. In contrast, standard DMD demonstrates the weakest performance.
	Turning to the prediction metrics in Figs.~\ref{fig:11}(b) and~\ref{fig:13}(b), it is observed that the KKF algorithm consistently demonstrates the poorest performance, while the Ens-DMD algorithm consistently outperforms all other methods. Notably, for the SSIM metric, the degradation in performance of the KKF algorithm is particularly pronounced. This indicates that the twin map structure produced by KKF can be significantly disrupted by noise.

\begin{figure}[t]
		\centering{}\includegraphics[width=3.5in]{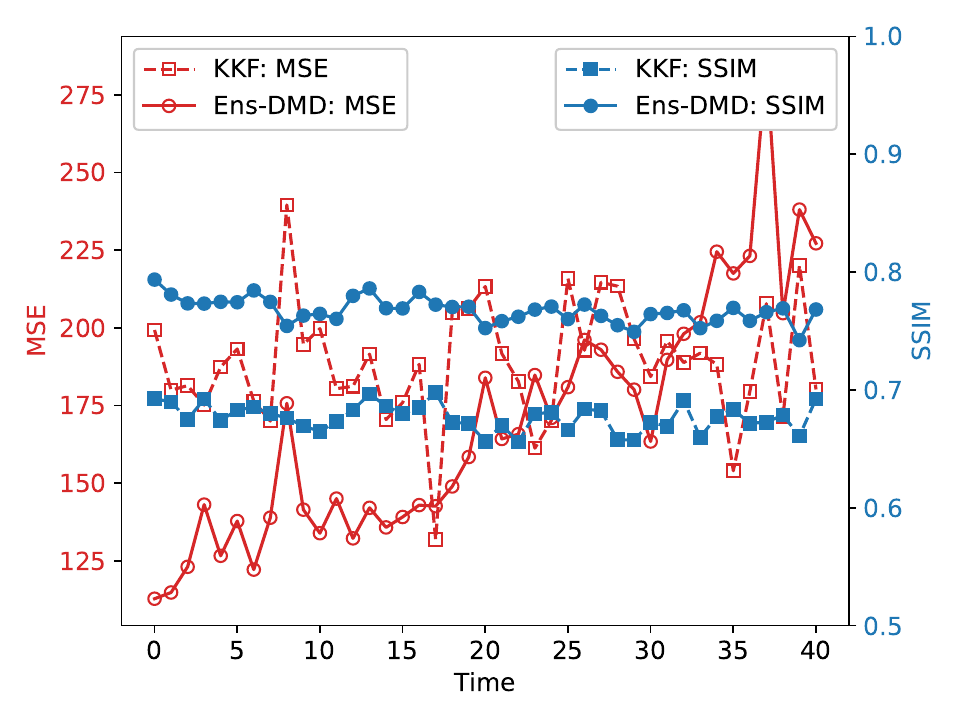}
		\caption{Prediction performance metric evolution.}
		\label{fig:14}
	\end{figure}

	We also investigate the predictive performance for future time steps in terms of MSE and SSIM metrics. As shown in Fig.~\ref{fig:14}, the results of all algorithms are averaged over 100 tests.
	The MSE performance of Ens-DMD experiences significant fluctuations only after predicting beyond 30 time steps. However, within the first 30 time steps, the MSE of Ens-DMD consistently remains lower than that of KKF. 
	Due to the inherent uncertainty of the Kriging method and the complexity of dynamic systems, the interaction between Kriging and Ens-DMD (or Kalman filtering) may lead to non-smooth behaviors as the time step increases. Additionally, it is observed that the SSIM performance of both algorithms shows a very slow decline, with DMD consistently outperforming KKF.

\begin{figure}[t]
		\centering
		\subfigure[]
		{\includegraphics[width=3.5in]{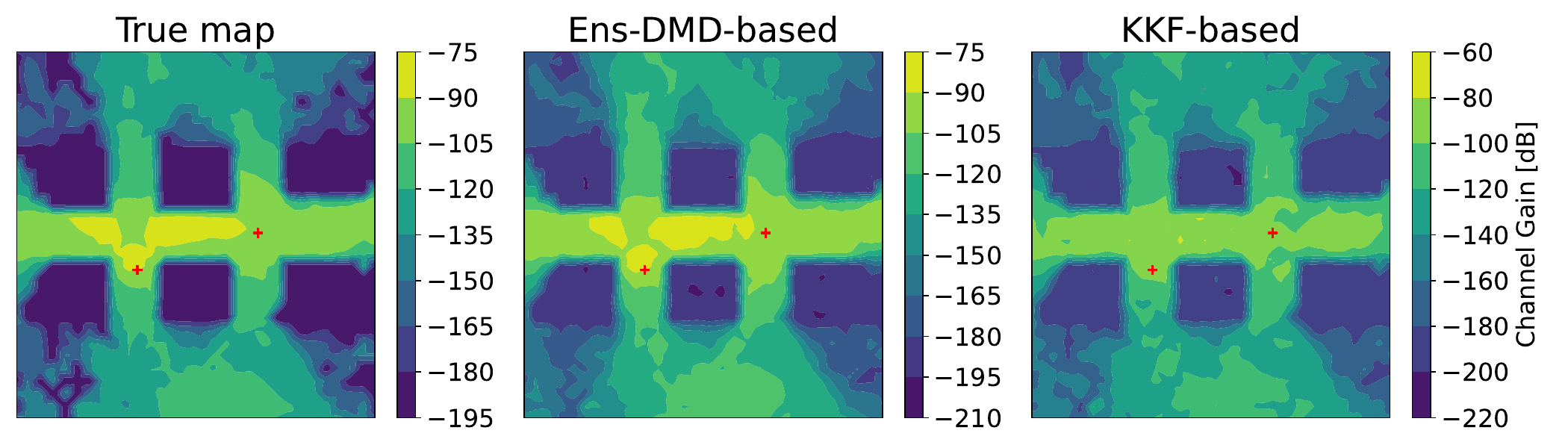}} 
		\subfigure[]
		{\includegraphics[width=3.5in]{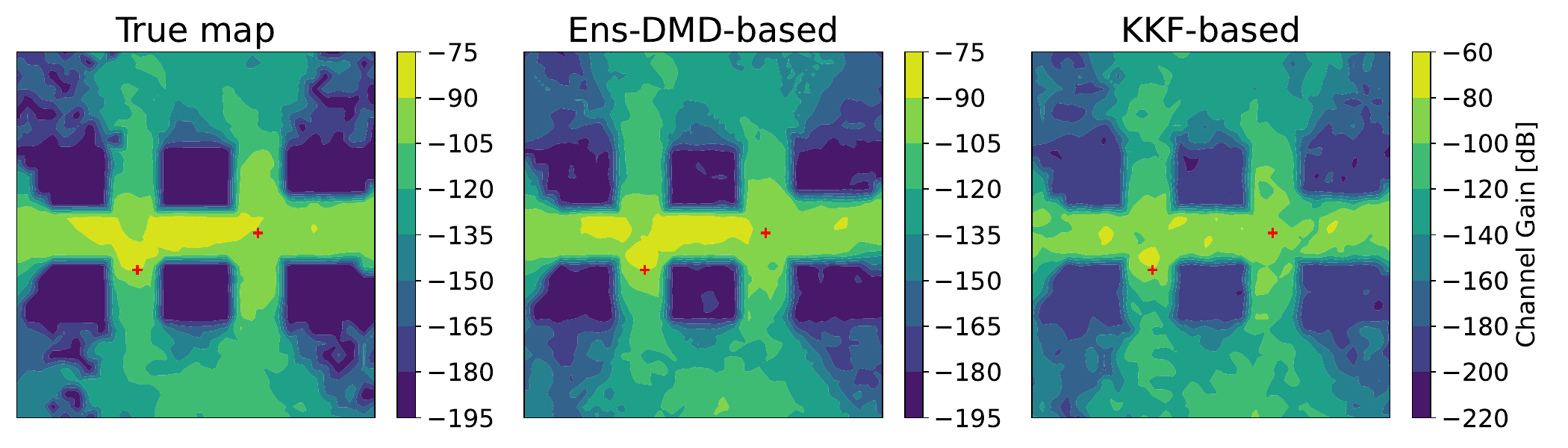}}
		\caption{Twin channel gain map with noisy dataset under the two-MD setup: (a) Reconstruction, and (b) prediction.}
		\label{fig:15}
	\end{figure}

    We consider a scenario with two MD nodes. As seen from Fig.~\ref{fig:15}, two mobile MDs are depicted by red plus signs.
	With noisy datasets, Fig.~\ref{fig:15}(a) compares the reconstructed gain maps using the Ens-DMD and KKF methods with ground truth. Both methods show reasonable alignment with the ground truth in low- to moderate-gain regions (e.g. between -160 dB and -80 dB). In high-gain regions (above -100 dB), KKF tends to overestimate the contour map due to its limited robustness against noise. In contrast, Ens-DMD demonstrates a slightly lower structural deviation in high-resolution reconstructions. Furthermore, based on the contour structure, it is difficult to locate the transmitters using the gain peak distribution from the KKF results.
	Fig.~\ref{fig:15}(b) illustrates the predicted radio contours. Ens-DMD achieves superior alignment with the true map across most gain levels, particularly in the dynamic region (above -120 dB). This highlights its robustness in capturing the dynamics of nonlinear systems. In contrast, KKF incurs a significant deviation at the gain peaks. In general, these results underscore the advantages of Ens-DMD in effectively handling noisy nonlinear channel environments.

	\subsection{Role of Twin Performance Metrics}\label{sec:4.4}
\begin{figure*}[t]
		\centering{}\includegraphics[width=6.5in]{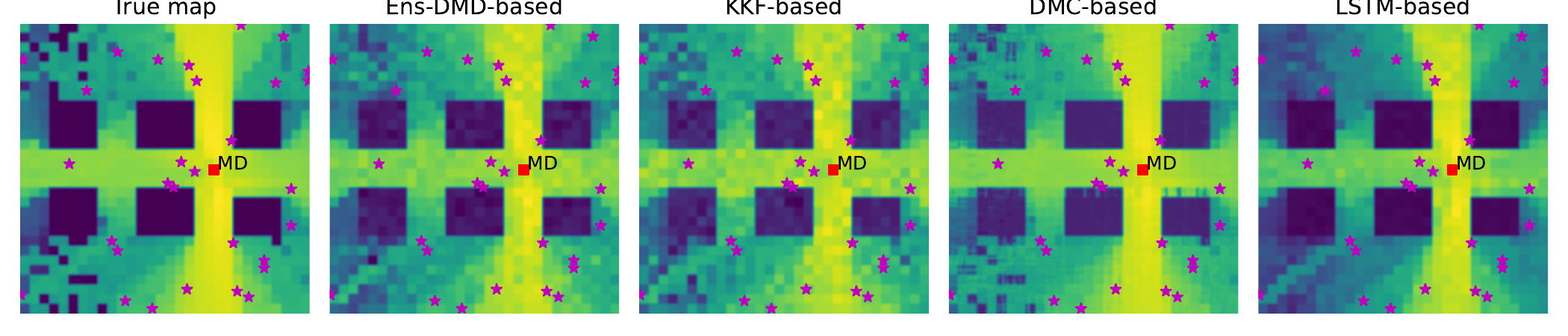}
		\caption{Simulation layout for radio map-based power allocation.}
		\label{fig:16}
	\end{figure*}

	Although the pixel-by-pixel error metrics are clear and important, SSIM is also a crucial metric that should not be overlooked in the context of wireless network applications. SSIM can effectively reflect the spatial distribution of channel gains.
	Such spatial distribution has been applied to the power allocation in the integrated sensing and communication system~\cite{10577661}.
	Therefore, we focus on a power allocation use case based on channel gain map twinning. Fig.~\ref{fig:16} visualizes the twin channel gain maps based on various algorithms, along with the random distribution of multiple users. Users are represented by purple pentagrams, while the MD is depicted by a red square.
	Since both the algorithms utilize the noisy dataset, the KKF-based channel gain map is scattered with random noise artifacts.
	
    % \begin{figure}[t]
    %         \centering
    %         \subfigure[]
    %         {\includegraphics[width=3.5in]{pics/fig13a.eps}} 
    %         \hspace{0.1in}
    %         \subfigure[]
    %         {\includegraphics[width=3.5in]{pics/fig13b.eps}}
    %         \caption{Radio map-based power allocation: (a) Simulation layout and (b) Achievable sum-rate vs. number of users with power allocation.}
    %         \label{fig:13}
    % \end{figure}
	
	\begin{figure}[t]
		\centering{}\includegraphics[width=3.5in]{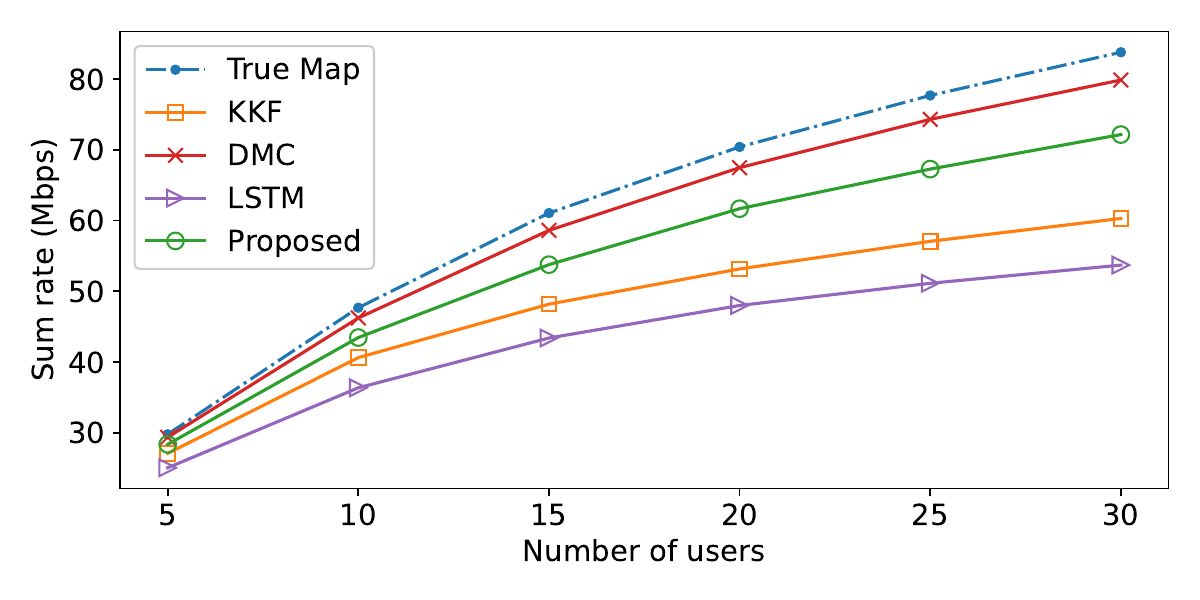}
		\caption{Achievable sum-rate vs. number of users with power allocation.}
		\label{fig:17}
	\end{figure}

	Fig.~\ref{fig:17} examines how channel gain maps produced by various algorithms affect the performance of the water-filling power allocation algorithm. The transmit power is set to 40 dBm and the received noise variance is -40 dBm~\cite{9127897}.	The signal bandwidth is set to 50 MHz. 
	Based on the grid points at which each user is located, channel gain values are retrieved from the channel gain maps obtained by various algorithms. Subsequently, water-filling power allocation is conducted using the channel gain information of all users, thus yielding the achievable sum rates based on the allocated power results. The outcomes depicted in Fig.~\ref{fig:17} represent averages over 1,000 trials.
	
	It is observed in Fig.~\ref{fig:17} that the achievable sum rates of all algorithms rise, as the number of users increases. However, the sum rate of the LSTM method remains the lowest, while the DMC algorithm outperforms the other algorithms and approaches the achievable sum rate derived from perfect channel gains. Nevertheless, the high running time of DMC may lead to outdated map twinning.
	The channel gain map generated by the LSTM yields a low SSIM when evaluated using the noisy dataset. This indicates that the proposed Ens-DMD algorithm can achieve a desirable balance between SSIM and computational delay.
	Higher user density correlates with an increased likelihood of users being located in noise artifact regions.
	Consequently, the discrepancies between power allocation results based on twinning maps and those derived from perfect channel gains become more significant. When there are over 10 users, the gap in achievable sum rates between algorithms widens progressively. This underscores the crucial role of SSIM in the context of communication applications.

	\section{Conclusions and Future Directions}\label{sec:5}
	In this paper, we developed a dynamic radio map twinning framework that works with small datasets. Leveraging the DMD principle, we modeled the radio map twinning process as a dynamical system.
	Considering the advantages and disadvantages of the DMD algorithm and its variants, we introduced the cDMD and eDMD algorithms to effectively extract coarse-grained and fine-grained evolution modes, respectively. To suppress the influence of noisy measurements, we proposed an Ens-DMD algorithm to effectively fuse the evolution modes through the designed median-based threshold mask technology.
	We demonstrated that the Ens-DMD algorithm can capture the underlying dominant evolution modes with much lower complexity than the KKF algorithm. We also investigated the reconstruction and prediction performance of our algorithm using four metrics. Simulation results validated the feasibility and effectiveness of the developed radio map twinning framework, which offers promising avenues for achieving equation-free, lightweight, and self-evolving channel twinning methods.
	
	The DMD-based twinning method, while effective in extracting temporal correlations between sequential map frames, does not fully exploit the intrinsic electromagnetic propagation characteristics in radio maps. As a future study, hybrid approaches that combine DMD with GAI and physics-inspired learning will be explored to further improve the channel twinning performance in complex environments. The dynamic modes extracted by DMD could provide physical priors for generative models to improve radio map synthesis, while embedding propagation physics into DMD models may enhance twinning robustness. In addition, current studies have relied primarily on centralized processing. For high-accuracy interpolation methods, such as matrix completion, distributed processing techniques can be studied to overcome the scalability challenges in dynamic radio map twinning.

	\bibliographystyle{IEEEtran}
	\bibliography{ref.bib}

\end{document}